\documentclass[acmlarge]{acmart}
\AtBeginDocument{%
  }

\setcopyright{cc}
\setcctype{by}
\acmJournal{IMWUT}
\acmYear{2025} \acmVolume{9} \acmNumber{3} \acmArticle{80} \acmMonth{9} \acmPrice{}\acmDOI{10.1145/3749517}

\acmISBN{978-1-4503-XXXX-X/18/06}
\usepackage{booktabs}
\usepackage{multirow}

\usepackage[utf8]{inputenc}
\usepackage{subfigure}
\usepackage{spverbatim}
\usepackage{booktabs}
\usepackage{array}
\usepackage{makecell}
\usepackage{longtable}
\usepackage{ragged2e}
\usepackage{tabularx} % 引入 tabularx 宏包
\usepackage{geometry}
\usepackage{enumitem}
\usepackage{listings}
\usepackage[normalem]{ulem}
\usepackage{arydshln}
\usepackage{array}
\newcolumntype{L}[1]{>{\raggedright\arraybackslash}p{#1}} % 可换行左对齐列

\usepackage[final,authormarkup=none,commandnameprefix=always]{changes}
\usepackage{xcolor}
\setdeletedmarkup{\textcolor{red}{\sout{#1}}}
\newcommand{\added}[1]{{\textcolor{black}{#1}}}

\begin{document}

%%
%% The "title" command has an optional parameter,
%% allowing the author to define a "short title" to be used in page headers.
%\title{From Conflict to Harmony: Developing Family Education Strategies through Parent-Child Conversations in Homework Involvement}
\title %[Exploring Emotions, Behaviours and  Conflicts Through Parent-Child Conversations in China]
{The Homework Wars: Exploring Emotions, Behaviours, and Conflicts in Parent-Child Homework Interactions
%Exploring Emotions, Behaviors, and Conflicts in Parental Homework Involvement through Parent-Child Conversations
}
%{Unpacking Parental Homework Involvement: Exploring Emotions, Behaviours and  Conflicts Through Parent-Child Conversations in China}

%% The "author" command and its associated commands are used to define
%% the authors and their affiliations.
%% Of note is the shared affiliation of the first two authors, and the
%% "authornote" and "authornotemark" commands
%% used to denote shared contribution to the research.
\author{Nan Gao}
%\authornote{Both authors contributed equally to this research.}
\email{nan.gao@nankai.edu.cn}
\orcid{0000-0002-9694-2689}
\affiliation{%
  \institution{Nankai University}
  \city{Tianjin}
  \country{China}
}
%\email{nangao@tsinghua.edu.cn}
\affiliation{%
  \institution{Tsinghua University}
  \city{Beijing}
  \country{China}
}
\affiliation{
  \institution{University of New South Wales}
  \city{Sydney}
  \country{Australia}
  \postcode{1466}
}

\author{Yibin Liu}
%\authornotemark[1]
\orcid{0009-0004-2925-7798}
\email{liuyibin@stumail.neu.edu.cn}
\affiliation{%
  \institution{College of Information Science and Engineering, Northeastern University}
  \city{Shenyang}
  \country{China}
}

\author{Xin Tang}
\orcid{0009-0009-9052-0897}
\email{tangxin@bupt.edu.cn}
\affiliation{%
  \institution{Beijing University of Post and Telecommunication}
  \city{Beijing}
  \country{China}
}

\author{Yanyan Liu}
\orcid{0009-0001-6929-6925}
\email{yy-liu22@mails.tsinghua.edu.cn}
\affiliation{%
  \institution{Institute of Education, Tsinghua University}
  \city{Beijing}
  \country{China}
}

\author{Chun Yu}
\orcid{0000-0003-2591-7993}
\email{chunyu@tsinghua.edu.cn}
\authornote{Corresponding author}
\affiliation{%
  \institution{Department of Computer Science and Technology, Tsinghua University}
  \city{Beijing}
  \country{China}
}

\author{Yun Huang}
\orcid{0000-0003-0399-8032}
\email{yunhuang@illinois.edu}
\affiliation{%
  \institution{School of Information Sciences, University of Illinois at Urbana-Champaign}
  \city{Champaign}
  \country{USA}
}

\author{Yuntao Wang}
\email{yuntaowang@tsinghua.edu.cn}
\orcid{0000-0002-4249-8893}
\affiliation{
  \institution{Department of Computer Science and Technology, Tsinghua University}
  \city{Beijing}
  \country{China}
  %\postcode{1466}
}

\author{Flora D. Salim}
\email{flora.salim@unsw.edu.au}
\orcid{0000-0002-1237-1664}
\affiliation{
  \institution{University of New South Wales}
  \city{Sydney}
  \country{Australia}
  \postcode{1466}
}

\author{Xuhai Xu}
\orcid{0000-0001-5930-3899}
\email{xx2489@columbia.edu}
\affiliation{%
  \institution{Google}
  \city{New York}
  \country{USA}
}

\author{Jun Wei}
\orcid{0000-0002-3468-0150}
\email{weijun8781@tsinghua.edu.cn}
\affiliation{%
  \institution{Institute of Education, Tsinghua University}
 \city{Beijing}
  \country{China}
}

\author{Yuanchun Shi}
\email{ shiyc@tsinghua.edu.cn}
\orcid{0000-0003-2273-6927}
\affiliation{%
  \institution{Department of Computer Science and Technology, Tsinghua University}
  \city{Beijing}
  \country{China}
}
\renewcommand{\shortauthors}{Gao et al.}

\begin{abstract}

Parental involvement in homework is a crucial aspect of family education, but it often triggers emotional strain and conflicts. Despite growing concern over its impact on family well-being, prior research has lacked access to fine-grained, real-time dynamics of these interactions. To bridge this gap, we present a framework that leverages naturalistic parent-child interaction data and large language models (LLMs) to analyse homework conversations at scale. In a four-week in situ study with 78 Chinese families, we collected 475 hours of audio recordings and accompanying daily surveys, capturing 602 homework sessions in everyday home settings. Our LLM-based pipeline reliably extracted and categorised parental behaviours and conflict patterns from transcribed conversations, achieving high agreement with expert annotations. The analysis revealed significant emotional shifts in parents before and after homework, 18 recurring parental behaviours and seven common conflict types, with \textit{Knowledge Conflict} being the most frequent. Notably, even well-intentioned behaviours were significantly positively correlated with specific conflicts. This work advances ubiquitous computing methods for studying complex family dynamics and offers empirical insights to enrich family education theory and inform more effective parenting strategies and interventions in the future.

\end{abstract}

%%
%% The code below is generated by the tool at http://dl.acm.org/ccs.cfm.
%% Please copy and paste the code instead of the example below.
%%

%%
%% Keywords. The author(s) should pick words that accurately describe
%% the work being presented. Separate the keywords with commas.

%\received{20 February 2007}
%\received[revised]{12 March 2009}
%\received[accepted]{5 June 2009}

%%
%% This command processes the author and affiliation and title
%% information and builds the first part of the formatted document.
\maketitle

\section{Introduction}

\begin{quote}
    \centering
   \textit{``Every night, millions of parents and kids shed blood, sweat and tears 
   \\over the kitchen table.''} - Sharon Begley \cite{begley1998homework}
\end{quote}

Family education is a cornerstone of individual development and societal progress, playing a pivotal role in shaping values, emotional intelligence, and lifelong learning. A critical component of family education is parental homework involvement, which significantly influences students' academic achievement, motivation, and well-being \cite{patall2008parent, dettmers2019antecedents}. Schools commonly encourage this involvement to support student success \cite{cooper1989synthesis} and strengthen parent-school connections \cite{heimgartner2012more}. In China, the intensity of parental engagement has increased notably, with average weekly homework involvement for primary school children increasing from 3.67 hours in 2010 to 5.88 hours in 2018 \cite{iccs}.

While parental homework involvement can foster academic progress and strengthen parent-child bonds, it can also be a source of conflict \cite{solomon2002helping, patall2008parent}. Many parents adopt ineffective strategies such as insufficient positive feedback \cite{pomerantz2007whom}, over involvement \cite{patall2008parent}, or unrealistic expectations \cite{pomerantz2007whom, cunha2015parents}. A 2019 survey of over 20,000 Chinese parents \cite{iccs} found that nine out of ten experienced emotional breakdowns during homework, and four out of ten exhibited out-of-control behaviours such as scolding or physical punishment. In extreme cases, homework-related stress triggered serious health outcomes, including heart attacks and strokes \cite{news_Mailonline_2020,news_Nov_Nov_2020}. This stress also affects children, potentially diminishing their learning interests, self-confidence, parent-child relationships, and long-term psychological health. Many experience intense frustration toward themselves, their parents, or their teachers, sometimes giving up on homework altogether \cite{xu2005homework}. Over time, repeated negative experiences can turn children off from homework or even lead to premature burnout \cite{corno2004homework}.

Despite the seriousness of these issues, the understanding of what actually happens during parental homework involvement remains limited. Previous research has largely relied on self-reports, interviews, or structured observations \cite{arrindell1999development, adair1984hawthorne}, which often fail to capture the nuances of emotional and behavioural dynamics unfolding in real time \cite{gan2019parental}. Without such detailed understanding, interventions remain generic, reactive, or inaccessible to families unaware of their own problematic patterns \cite{mcewan2004deal, waters2020keep}.
Moreover, existing LLM-based studies rarely addressed spontaneous, emotionally laden conversations in natural family settings, leaving a critical gap in modelling moment-by-moment dynamics that are essential for designing culturally responsive and context-aware interventions.
%how specific behaviours and conflicts unfold moment-by-moment. Without nuanced modelling of such situational human interaction data, we lack the foundation for culturally sensitive, context-aware interventions—motivating our present study.

%Despite the seriousness of these issues, we still lack a detailed understanding of the emotional and behavioral dynamics that unfold during homework sessions. Prior studies have largely relied on retrospective methods—surveys, interviews, or structured observations \cite{arrindell1999development, adair1984hawthorne}—which often fail to capture the nuances of real-time interaction \cite{gan2019parental}. As a result, many interventions remain generic, reactive, or inaccessible to families unaware of their own problematic patterns \cite{mcewan2004deal, waters2020keep}. Moreover, existing LLM-based studies rarely examine spontaneous, emotionally charged conversations in natural family settings, leaving a critical gap in modeling moment-by-moment dynamics that are essential for designing culturally responsive and context-aware interventions.

To address this gap, we conducted a 4-week in situ study with 78 Chinese families, collecting 602 valid audio recordings of homework interactions and accompanying survey data. We propose leveraging large language models (LLMs) to analyse these real-world conversations, aiming to extract and categorise patterns of parental behaviours and conflicts. We further validate the reliability of these automated analyses by comparing them against annotations from four human experts. Specifically, we ask:
\textit{1. What does parental homework involvement in China look like, and do parents experience emotional changes post-homework?}
\textit{2. What types of parental behaviours and parent-child conflicts arise during homework involvement in China?}
\textit{3. How are parents' emotions, behaviours, and conflicts interconnected during homework involvement?}

Our findings reveal significant shifts in parents’ emotional states, measured by the \textit{PAD} (pleasure, arousal, dominance) model \cite{bradley1994measuring} before and after homework, with all dimensions showing statistical significance (p < 0.001).
Using LLM-based analysis and expert input, we identified 18 common parental behaviours (categorised as positive, neutral, or negative) and 7 types of conflict, with \textit{Knowledge Conflict} being the most common. Interestingly, even well-intended behaviours like \textit{Unlabelled Praise} were significantly associated with specific conflicts, such as \textit{Expectation Conflict}, highlighting the nuanced nature of interaction patterns. 
In sum, our paper makes the following contributions:

\begin{itemize}
    \item We conducted a 4-week in situ study with 78 Chinese families, collecting 602 audio recordings (475 hours) and daily surveys on real-world parental homework involvement. From this data, we identified 18 parental behaviours (6 positive, 6 neutral, and 6 negative) and 7 common parent-child conflict types specific to the Chinese context.

    \item We developed and validated an LLM-based framework for extracting parental behaviours and conflicts from extensive conversations, achieving high agreement with four human expert annotations, demonstrating the feasibility of automated analysis in complex home environments.

    \item We analysed the emotional, behavioural, and conflict dynamics during parental homework involvement, revealing several key findings. For instance, even positive and neutral parental behaviours were significantly linked to certain conflicts. 
    These insights deepen the theoretical understanding of family education and inform more effective parenting strategies and interventions in the future.

\end{itemize}

\section{Related Works}

\subsection{Parental Homework Involvement in Education}

Parental homework involvement refers to \textit{parents' monitoring, supervision, and participation in their children's schoolwork and academic performance} \cite{pomerantz2007whom}. This involvement is widely recognised as a key factor in children's academic success, motivation, and well-being \cite{patall2008parent, dettmers2019antecedents, cooper1989synthesis}. While much of the literature has focused on Western contexts, studies in China reveal distinct cultural patterns. Chinese parents often take a more hands-on approach, using strategies such as reasoning, monitoring, and even criticism, reflecting a broader societal emphasis on academic excellence and a strong sense of parental duty \cite{kim2013parents}. Gan et al. \cite{gan2019parental} have identified a spectrum of involvement in China, from supportive to disengaged, with supportive approaches most strongly associated with academic success \cite{gan2019parental}. These findings underscore the need to understand homework involvement within diverse cultural settings.

Parental homework involvement has typically been studied through self-report surveys  (e.g., EMBU \cite{arrindell1999development}, QPH \cite{dumont2014quality}), interviews, and observations. While these methods provide valuable insights, they often overlook the emotional and behavioural nuances of real-world interactions. Observations may be affected by the \textit{Hawthorne Effect} \cite{adair1984hawthorne}, and self-reports can introduce bias \cite{gao2021investigating} and may not fully reflect the subtleties of everyday involvement. Broad typologies, such as Gan et al.’s four categories of involvement \cite{gan2019parental}, may fail to capture the complexity of everyday parent–child dynamics during homework. As a result, our understanding of what truly occurs during real-world homework interactions remains incomplete.

\subsection{Emotional Experiences and Parent-Child Conflicts During Homework Involvement}

While parental homework involvement is often linked to positive educational outcomes, it can also lead to emotional strain and conflicts within families. Nnamani et al. \cite{nnamani2020impact} highlighted that while such involvement supports children's emotional adjustment and performance, the emotional burden on parents is frequently overlooked, especially in cultures like China where academic achievement is highly prioritised.
Parents often struggle to balance fostering autonomy with maintaining control over their children's learning, a tension that can produce stress for both parties \cite{cunha2015parents}. This emotional strain can create a feedback loop, where parental stress negatively affects children, escalating conflict during homework sessions \cite{moe2018brief}. \added{Kim et al. \cite{kim2020dyadic} addressed this issue with \textit{Dyadic Mirror}, a wearable smart mirror that reflects parents’ facial expressions from their child’s perspective, aiming to raise awareness of emotional dynamics during interactions.}

Conflicts tend to intensify during adolescence, when academic pressure and developmental pushback collide.  Solomon et al. \cite{solomon2002helping} explored how homework involvement can become a source of conflict and found that the pressure parents feel to ensure their children's academic success can exacerbate tensions, often turning homework sessions into battlegrounds where unresolved issues about control and expectations surface. Similarly, Suarez et al. \cite{suarez2022parental} observed heightened family conflict during the COVID-19 pandemic, as increased parental involvement in home-based learning amplified stress.%This finding is echoed by Suarez et al. \cite{suarez2022parental}, who reported high levels of family conflict and stress during the COVID-19 pandemic, a time when parental involvement in homework increased dramatically due to school closures and remote learning. 

These findings reveal the complex emotional terrain of homework involvement. 
%Well-meaning efforts to support learning may unintentionally lead to tension and conflict. 
\added{ However, most existing studies do not capture the unfolding emotional and interactional dynamics in the moment, leaving a gap in our understanding of how behaviours, emotions, and conflicts are interconnected in practice.}

\subsection{\added{LLM-Assisted Qualitative Coding}}

Recent studies have begun exploring how LLMs can support traditional qualitative coding. 
For example, Xiao et al. \cite{xiao2023supporting} used GPT-3 with expert-developed codebooks for deductive coding, finding fair to substantial agreement with human coders and emphasising the importance of prompt design. Dai et al. \cite{dai2023llmintheloopleveraginglargelanguage} proposed an \textit{LLM-in-the-loop} framework using GPT‑3.5, achieving coding quality comparable to human analysts while significantly reducing manual effort. Dunivin \cite{dunivin2025scaling} advanced a hybrid pipeline that incorporates iterative codebook refinement and chain-of-thought prompting to align model outputs with human interpretive norms.
Empirical evidence increasingly suggests that LLMs can approach or even surpass human inter-coder reliability for certain coding dimensions, such as sentiment and discourse structure \cite{bojic2025comparing}. 

However, existing studies still struggle with nuanced cues like sarcasm and emotional intensity. In applied contexts, findings remain promising but mixed. Liu et al. \cite{liu2024llm} evaluated various prompting strategies for coding virtual tutoring transcripts, while Liu et al. \cite{liu2025qualitative} reported that LLM annotations of educational dialogues closely aligned with expert labels, although the most challenging constructs were those with low human agreement as well. In a multilingual context, Tornberg et al. \cite{tornberg2024large} found that instruction-tuned LLMs outperformed both traditional classifiers and human coders in identifying political stance from tweets, particularly when interpreting implicit or context-dependent meanings.

Despite these advances, most prior studies focus on structured or semi-structured texts such as interviews, chat logs, or social media posts. For example, Aldeen et al.~\cite{aldeen2024end} analysed user reviews to surface problematic behaviours in Alexa skills. Few works engage with real-world, unstructured conversational data, and even fewer examine emotionally charged family interactions. Additionally, existing frameworks are predominantly developed in Western contexts, limiting their generalizability to cultures with different linguistic norms, family roles, and emotional expressions.

In contrast to prior work, our research distinguishes itself in several ways: (1) 
We focus on a culturally distinct, high-stakes context: Chinese parent-child homework interactions, where academic expectations and emotional tensions are uniquely intense  \cite{kim2013parents, suarez2022parental};
(2) we move beyond structured or retrospective data by analysing naturalistic, unstructured audio recordings of in situ family conversations-capturing the emotional immediacy and behavioural complexity often missed in surveys or interviews;
(3) we bridge behavioural, emotional, and relational layers in a unified analytical framework, combining expert-level interpretive depth with the scalability of LLMs. Our approach expands the frontier of computational qualitative research into emotionally charged, culturally embedded, and interactionally rich real-world domains.

% Our study distinguishes itself in several ways: (1) we focus on the Chinese cultural context, shaped by unique parental expectations and pressures \cite{kim2013parents, suarez2022parental}; (2) unlike prior work that largely depends on subjective data, we utilize audio recordings of real-world homework sessions for a richer and more objective analysis; (3) by exploring the interplay of parental behaviours, emotions, and conflicts, we aim to deepen understanding of the complexities in homework involvement, contributing valuable insights to the family education research and designing technologies to improving parenting practices in China.

\section{Understanding Parental Homework Involvement in China}

%To design technologies for improving parental practices, it's crucial to understand parental involvement in homework scenarios. 
In this research, we conducted a 4-week real-world data collection, recording parent-child conversations during homework sessions. 
%This approach provides detailed insights necessary for designing technologies for improving parenting strategies. %Specifically, we aim to answer the following questions: \textit{1. What does parental homework involvement in China entail, and do parents experience emotional changes post-homework?}
%\textit{2. What types of parental behaviours and parent-child conflicts arise during homework involvement in China?}
%\textit{3. How are parents' emotions, behaviours, and conflicts interconnected during homework involvement?} 
The data collection was approved by the Human Research Ethics Committee at our university.

\begin{table}

%\begin{minipage}[]{6cm}
\caption{Demographics of parent participants.}
\label{tab:par_parent}
\begin{tabular}{@{}llll@{}}
\toprule
\textbf{Item}               & \textbf{Option    }                             & \textbf{Count} & \textbf{Percentage} \\ \midrule
\multirow{2}{*}{\textit{Gender}}    & Female (mother)                        & 68    & 87.18\%    \\
                           & Male (father)                          & 10    & 12.82\%    \\\hline
\multirow{4}{*}{\textit{Age}}       & $\leq$ 30                                    & 5     & 6.41\%     \\
                           & 31-40                                  & 48    & 61.54\%    \\
                           & 41-50                                  & 25    & 32.05\%    \\
                           & $\geq$50                                    & 0     & 0.00\%     \\\hline
\multirow{5}{*}{\begin{tabular}[c]{@{}l@{}}\textit{Education} \\ \textit{Level}\end{tabular}} & Junior high school and below           & 0     & 0          \\
                           & High school/Secondary school & 4     & 5.13\%     \\
                           & Associate degree (Junior college)      & 7     & 8.97\%     \\
                           & Bachelor's degree                      & 45    & 57.69\%    \\
                           & Postgraduate degree and above          & 22    & 28.21\%    \\ \bottomrule
\end{tabular}
%\end{minipage}

\iffalse
\hspace{2cm}
\begin{minipage}[]{6cm}
\caption{Basic information of the children of parent participants.}
\label{tab:par_child}
\begin{tabular}{@{}llll@{}}
\toprule
\textbf{Item}               & \textbf{Option}            & \textbf{Ct.} & \textbf{Percent.} \\ \midrule
\multirow{2}{*}{\textit{Gender}}& Female (daughter) & 38    & 48.72   \\
& Male (son)        & 40    & 51.28\% \\\hline
\multirow{3}{*}{\textit{Grade}} & Year 1            & 31    & 39.74\% \\
& Year 2            & 27    & 34.62\% \\
& Year 3            & 20    & 25.64\% \\\hline
\multirow{5}{*}{\begin{tabular}[c]{@{}l@{}}\textit{Academic} \\ \textit{Rank in} \\ \textit{Class}\end{tabular}} & Bottom            & 1     & 1.28\%  \\
& Below Average     & 9     & 11.54\% \\
& Average           & 13    & 16.67\% \\
& Above Average     & 35    & 44.87\% \\
& Top               & 20    & 25.64\% \\ \bottomrule 
\end{tabular}

\end{minipage}
\fi
\end{table}

\subsection{Participants}
\label{subsec: participants}
Participants were recruited based on specific criteria: they had to be parents of primary school students in Years 1-3, who were regularly assisted with homework and communicated in Mandarin at home. We distributed advertising leaflets and posters through various channels, including word of mouth, social platforms like WeChat, and school principals who facilitated further distribution through teachers. 

Within a week of registration, a total of 121 individuals completed the registration questionnaire and provided consent for data collection. Among them, 106 met the eligibility criteria, with 15 participants excluded due to reasons such as their children being in Year 4, the involvement of home-based teachers rather than parents, or infrequent homework involvement (defined as one or fewer instances per week). Out of the 106 eligible participants, 85 were successfully contacted; however, seven of them withdrew without contributing data.

The final sample consisted of 78 Chinese parents, with only one parent from each family included.  Detailed demographic information is provided in Table \ref{tab:par_parent}. Notably, the majority of participants (87.18\%) were mothers. Additionally, the education level of the participants was generally higher than the national average in China \cite{gps}. This may be attributed to two primary factors: 1) more educated parents might be more motivated to participate in obtaining a family education report as part of the compensation; 2) these parents likely had better access to our advertisements through social media and other channels. %Regarding the children, their academic rankings, as reported by parents, were not evenly distributed, with most being described as \textit{``above-average''}. This imbalance suggests a sampling bias, where parents of higher-achieving students might be more invested in education and thus more inclined to participate. 
We acknowledge this limitation and address it further in Section \ref{sec:limit}.

\subsection{Procedure}

Before formal data collection began, participants completed a registration questionnaire that gathered basic demographic information, details on their habits of parental home involvement, and their satisfaction and emotions related to these activities. The insights gained from this preliminary stage informed the design of daily surveys. 

During the data collection, participants first completed an online background survey\footnote{The first background survey data was not used in this research.}. They were then instructed to record audio during homework sessions with their children based on their usual routines (not necessarily daily). To minimise the \textit{Hawthorne Effect} \cite{adair1984hawthorne} — where individuals alter their behaviour due to awareness of being observed - participants were encouraged to act naturally and were assured that their privacy would be strictly protected, with data used solely for research purposes.

After recording, participants could upload their audio via a private link or send it to our assistant, who stored it on secured hardware. On days when audio was recorded, participants were asked to complete a daily online questionnaire immediately after the homework session. Each parent could submit up to 10 sets of audio recordings and corresponding daily questionnaires over the four weeks. This flexibility accommodated variations in the frequency of homework involvement in different families, with some parents participating daily and others two or three times per week. Additionally, participants were asked to complete a second background survey\footnote{The second background survey data was not used in this research.} at the end of the first week. To minimise the burden, parents used their own smartphones or preferred devices to record the entire homework process. Although some parents may check in only at the end of the session, we still require recordings from start to finish to fully understand when and how parents intervene during homework sessions. 

%Participants could upload audio via a private link or send it to our assistant, who stored it on secured hardware. On days when audio was recorded, participants completed a daily online questionnaire immediately after the session. Each parent could submit up to 10 sets of audio recordings and daily questionnaires over the four-week period, allowing for variation in homework involvement frequency, with some parents participating daily and others two to three times per week. Additionally, participants completed a second background survey\footnote{The second background survey data was not used in this research.} at the end of the first week. To minimize burden, parents used their own devices to record the entire homework process, even if they usually only check in at the end, allowing us to capture how and when parents intervene during homework sessions.

\begin{figure}
    \centering
    \begin{minipage}[t][6cm]{0.45\textwidth}
        \includegraphics[width=0.95\textwidth]{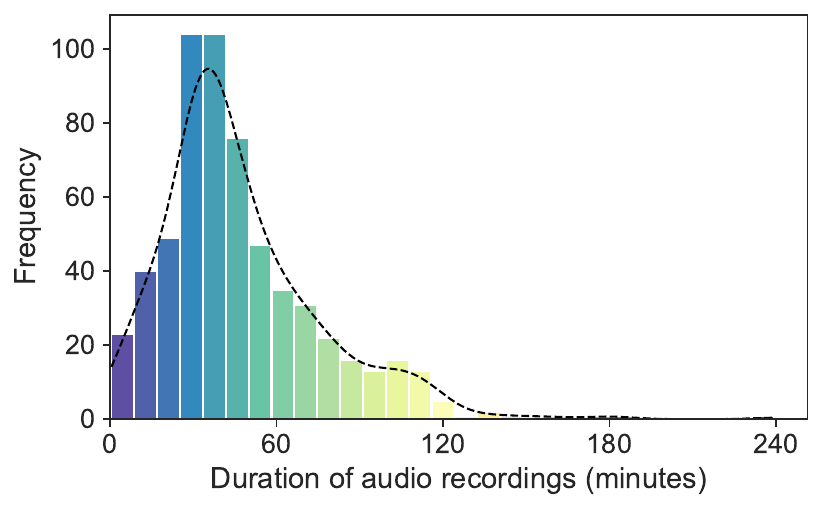}
        \caption{The distribution of audio recording durations.}
        \label{fig:dis_length}
    \end{minipage}
    \hspace{0.2cm}
    \begin{minipage}[t][6cm]{0.45\textwidth}
    \centering
        \includegraphics[width=0.94\textwidth]{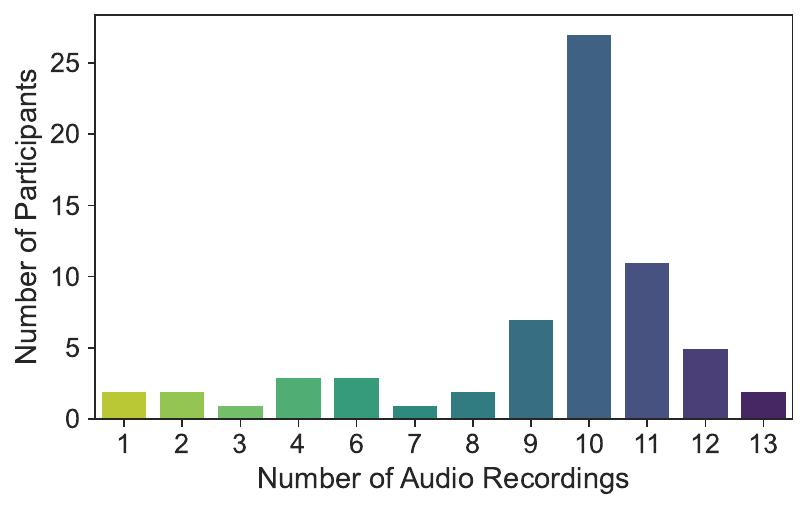}
        \caption{Number of audio recordings for different participants.}
        \label{fig:dis_count}
    \end{minipage}
    
\end{figure}
Participants were compensated up to RMB 240 for their involvement (RMB 40 for the background survey and RMB 20 for each audio recording and its corresponding daily survey, up to a maximum of ten sets). Additionally, those who provided complete data would receive a family education report with tailored suggestions from our team of family education specialists. This incentive was designed, in part, to encourage parents to behave as naturally as they would at home. It is worth noting that the participation in this research project is entirely voluntary. Participants were free to withdraw from the study at any time. To ensure privacy, all participants were anonymised and assigned a unique ID during the data collection.

%It is worth noting that participation in this research project isvoluntary. Participants are free to withdraw from the project at any stage if they change their minds. Besides, we anonymized all the participants to protect their privacy

\subsection{Collected Data}

We gathered data from several sources: registration questionnaires, background surveys, audio recordings of parental homework involvement, and daily surveys. We collected 121 registration questionnaires and 78 background surveys, but only used the data for participants who completed both. We also collected 602 valid audio recordings and 620 daily surveys. 

For the record data, while we only require one complete audio file, some parents accidentally produce multiple recordings due to phone calls or pressing the wrong button. These recordings, which constitute about 10.43\% (66 out of 633 homework sessions) of our data, are merged into a single file before processing, leaving out 603 audio files. After removing one corrupted file, we collected 602 valid audio files. The total duration of these recordings is 474.89 hours, with an average length of 47.33 minutes per audio. Figure \ref{fig:dis_length} shows the distribution of audio recording durations, while Figure \ref{fig:dis_count} illustrates the number of recordings contributed by different participants. On average, 24 recordings were submitted daily, with 66 parents successfully contributing at least one recording. 

\begin{figure}
    \subfigure[ {Pleasure} \label{subfig: age}]{\includegraphics[width=0.285\textwidth]{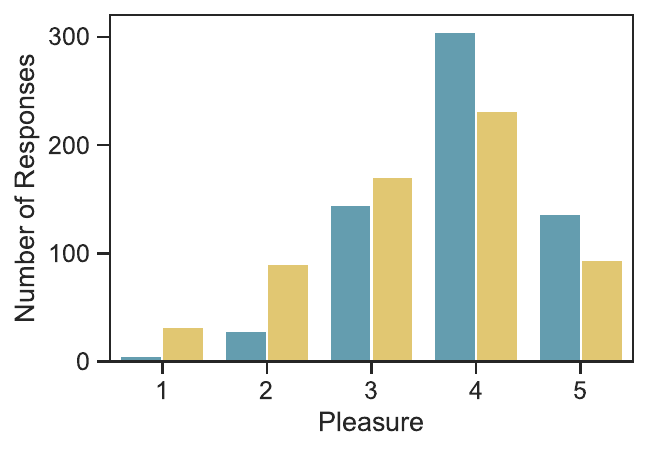}}
	\hspace{0cm}
    \subfigure[ {Arousal}\label{subfig: venn}]{\includegraphics[width=0.285\textwidth]{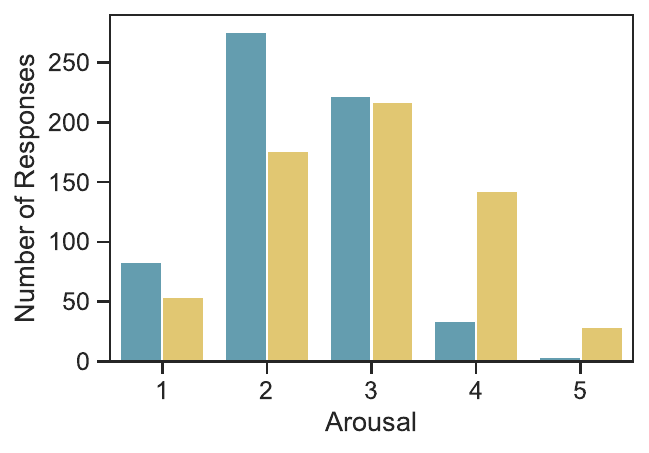}}
    \subfigure[ {Dominance}\label{subfig: venn}]{\includegraphics[width=0.405\textwidth]{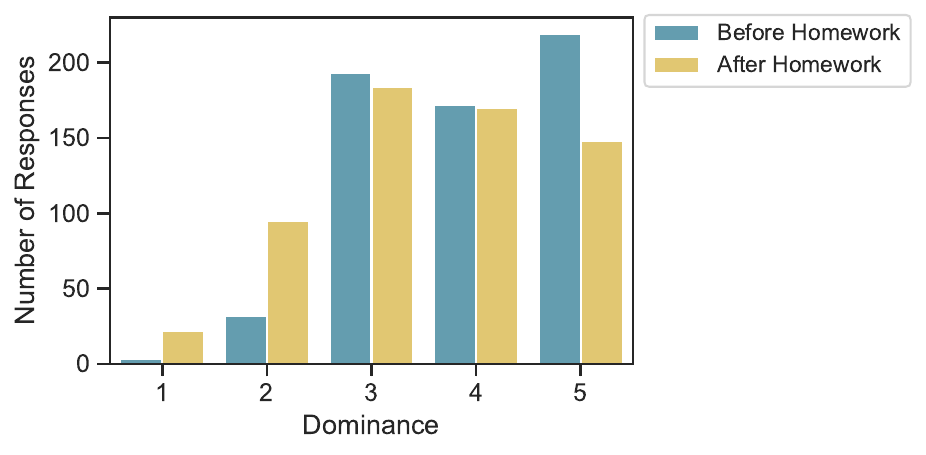}}
    \caption{Distribution of parental emotions before and after homework involvement.}
    \label{fig:self-report dis}
\end{figure}

For the daily survey\footnote{Additional data, such as child behaviour evaluations, parental behaviours, and daily stress, were collected but not utilized in this research.}, it assessed the parents' affective reactions before and after homework involvement, focusing on pleasure, arousal, and dominance, using the \textit{Self-Assessment Manikin} (SAM) \cite{bradley1994measuring} and \textit{PAD} emotion state model \cite{mehrabian1974approach}. Figure \ref{fig:self-report dis} presents the parents' perceived emotions before and after homework, with pleasure (1-5) ranging from extremely unhappy to extremely happy, arousal (1-5) from completely calm to highly aroused, and dominance (1-5) from highly submissive to highly dominant.

%To capture the overall trend and the diversity of emotional responses among parents, we conducted both group-level and individual-level analyse to thoroughly understand the emotional experiences of parents before and after homework involvement, based on self-reported data. First, we calculated the mean emotional values for each parent, both before and after homework. Then, we conducted a paired sample t-test on these group mean values, comparing the pre-homework means to the post-homework means. We observed statistically significant differences in the mean values across all dimensions, with pleasure (p < 0.001), arousal (p < 0.001), and dominance (p < 0.001) all showing significant changes. It revealed whether parents, as a group, experienced a statistically significant emotional shift due to homework involvement. Next, we examined whether individual parents showed significant changes in their emotions before and after homework. This allowed us to identify those parents who experienced notable emotional shifts, providing insight into the variability of responses.

\subsection{Data Processing}
\subsubsection{Audio Preprocessing}
The audio preprocessing followed three key steps:
(1) \textit{Conversion to WAV Format}. 
%All audio files were converted to WAV, a lossless format that preserves the full quality of the recordings. This was essential to ensure no audio information was lost, as maintaining high fidelity was crucial for accurate analysis. 
All audio files were converted to lossless WAV to preserve full recording quality, ensuring no information was lost for accurate analysis.
(2) \textit{Resampling}. 
%The audio was resampled to a uniform rate of 16 kHz, a standard in speech processing. This rate strikes a balance between maintaining sufficient detail for accurate speech recognition and managing computational efficiency.
Audio was resampled to 16 kHz, balancing sufficient detail for speech recognition with computational efficiency.
(3) \textit{Normalization}.  Audio levels were normalised to ensure consistent volume across recordings, preventing bias during feature extraction. Noise removal was deliberately avoided after preliminary tests showed it could distort or remove crucial elements, particularly the child’s voice.

\subsubsection{Transcription}

Transcription was performed on the normalised audio files, preserving periods of silence to retain the natural flow of conversation. We employed the Xunfei API \cite{xunfei} for automatic transcription, applying a "2-second rule" for utterance segmentation: pauses longer than two seconds were treated as meaningful breaks and used to divide the conversation into distinct turns. Given the informal nature of parent-child conversations, automatic transcription often introduces errors such as homophone confusion, misinterpretations, and segmentation issues. To address this, we used a custom-designed prompt (see Appendix \ref{app:prompt_transcription}) with the GPT-4o model for error correction. This prompt was tailored to guide the model in correcting inaccuracies while preserving the original tone and flow of the dialogue. Our guiding principles were minimal intervention, conversational coherence, and transcription accuracy. All survey questions and transcripts were originally conducted in Mandarin Chinese. Consequently, both the coding prompts and the codebook used in qualitative analysis were also written in Mandarin to preserve linguistic nuances and ensure faithful interpretation. For clarity and accessibility to an international audience, we translated the main codebook (Table~\ref{tab:behaviours} and Table \ref{tab:conflict}) and prompt templates (Appendix \ref{app:prompt_transcription}) into English. These translations aim to convey the original meanings while enabling non-Chinese-speaking readers to understand the analytical process.

%Our guiding principles were minimal intervention, conversational coherence, and transcription accuracy. All interviews and transcripts were originally conducted in Mandarin Chinese. Accordingly, the coding prompts and codebook used in the qualitative analysis were also written in Mandarin to retain linguistic nuances and ensure faithful interpretation. For clarity and accessibility to an international audience, we translated the main codebook (Tables~\ref{tab:behaviours}, \ref{tab:conflict}) and prompt templates (Appendix~\ref{app:prompt_transcription}) into English. These translations aim to convey the original meanings while enabling non-Chinese-speaking readers to understand the analytical process.

\subsubsection{Role Recognition} \label{subsec: roles}
The Xunfei API could identify multiple speakers but labelled them generically (e.g., "Speaker 1"). This posed challenges in identifying specific roles, such as distinguishing between parent and child or between different speakers assigned the same label. To resolve this, we designed a model-specific prompt (see Appendix \ref{app: prompt_role recognition}) to clarify speaker roles. The prompt used contextual cues and conversational patterns to assign accurate roles to each speaker. It also addressed cases where speakers with the same label might actually represent different individuals, ensuring accurate identification of all participants.

%The Xunfei API used in transcription could identify multiple speakers but only labeled them generically as "Speaker 1," "Speaker 2," etc., without specifying their identities or roles. This presented two challenges: (1) determining which speaker corresponded to which role (e.g., parent or child) and (2) identifying whether multiple speakers of the same label belonged to the same role or different ones. To address these challenges, we developed a specific prompt to clarify speaker roles post-transcription. The prompt was designed to instruct the model to distinguish between different speakers based on contextual cues and conversational patterns, ensuring accurate role identification. The prompt also guided the model in determining whether speakers labeled identically by the API actually represented different roles.

%\tobe{Expert Labeling: We selected 30 audio recordings for a thorough evaluation. Two experts independently labeled the roles based on the transcription. If their labels matched, the roles were confirmed. In cases of disagreement, the experts listened to the audio to resolve the discrepancies and accurately define the roles. These expert-labeled roles were considered the ground truth. Model Testing: For each of the 30 audio recordings, the role recognition prompt was run 10 times to assess consistency and accuracy. The results were compared against the expert-labeled ground truth to evaluate the effectiveness of the role recognition process.}

\section{Understanding Parents' Emotion Dynamics During Homework Involvement}

%Understanding how parents’ emotions evolve during homework involvement is key to supporting healthier parent-child interactions. 
In this section, we first analyse self-reported data to identify overall emotional shifts before and after sessions. Then, we use fine-grained emotion annotations from transcribed interactions to capture moment-to-moment emotional fluctuations.

\subsection{Perceived Emotion Shifts after Homework Involvement} 
To understand parents' emotional experiences before and after homework involvement, we conducted both group-level and individual-level analyses using self-reported data. For the group-level analysis, we calculated the mean emotional values for each parent before and after homework, followed by a paired sample t-test to compare the pre- and post-homework means across the entire group. This revealed significant \added{emotional shifts in all dimensions}-valence (p < 0.001), arousal (p < 0.001), and dominance (p < 0.001). 
Specifically, parents experienced a decrease in pleasure, a decrease in arousal, and a reduction in their sense of control (dominance) after homework involvement.

\iffalse
\begin{figure}
    \centering
    \includegraphics[width=0.99\linewidth]{figure/significance_heatmap.pdf}
    \caption{Significance heatmap of pleasure, arousal, and dominance. Red, blue, and green colours indicate participants with statistically significant differences (p < 0.05) in pleasure, arousal, and dominance, respectively, before and after homework involvement.}
    \label{fig:heatmap}
\end{figure}
\fi
\begin{figure}
    \subfigure[ {Pleasure} \label{subfig: age}]{\includegraphics[width=0.75\textwidth]{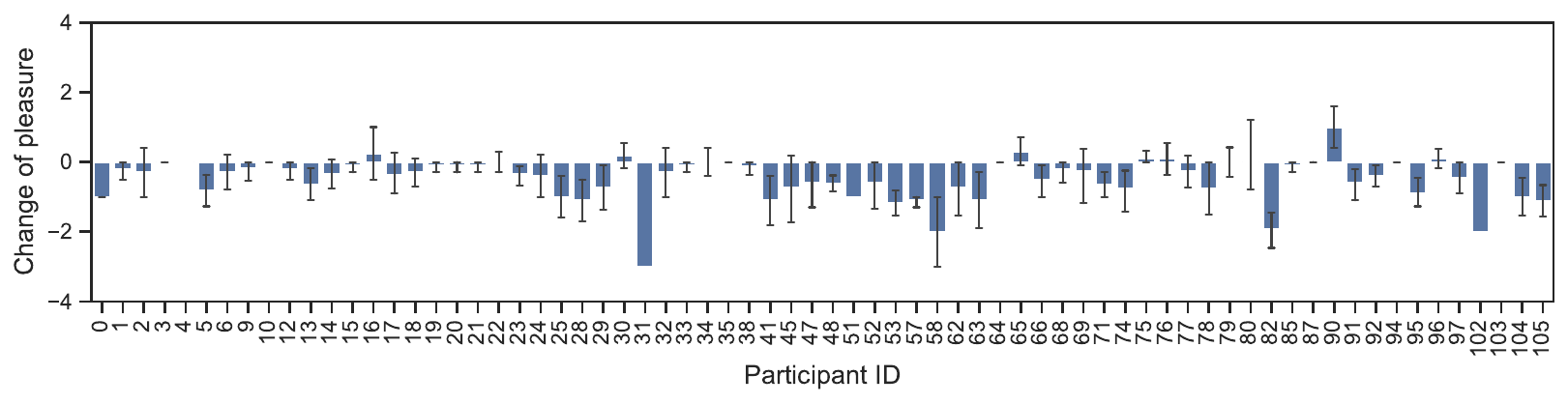}}
	\hspace{0cm}
    \subfigure[ {Arousal}\label{subfig: venn}]{\includegraphics[width=0.75\textwidth]{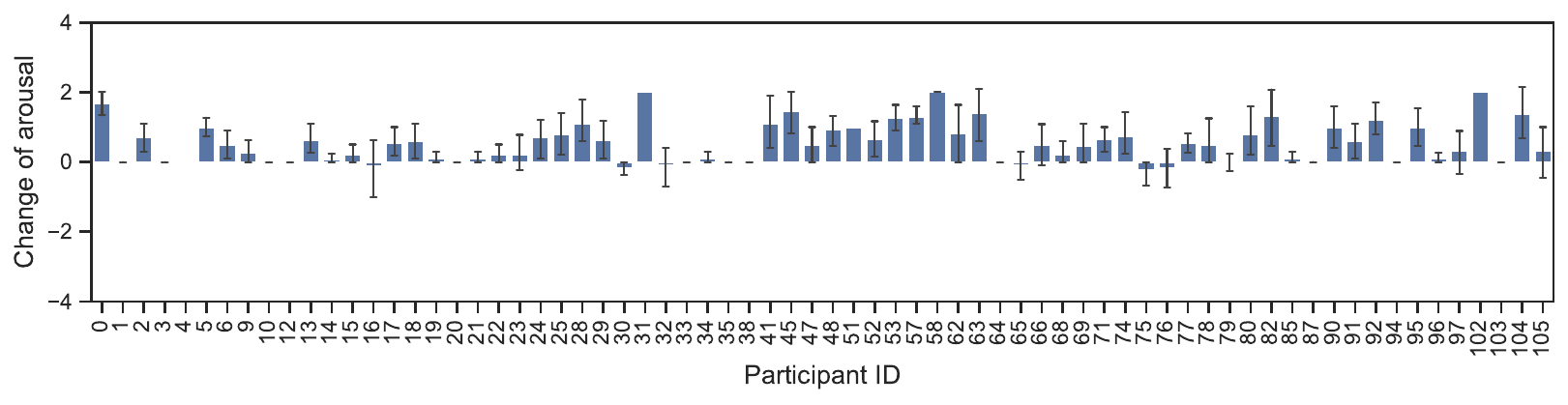}}
    \subfigure[ {Dominance}\label{subfig: venn}]{\includegraphics[width=0.75\textwidth]{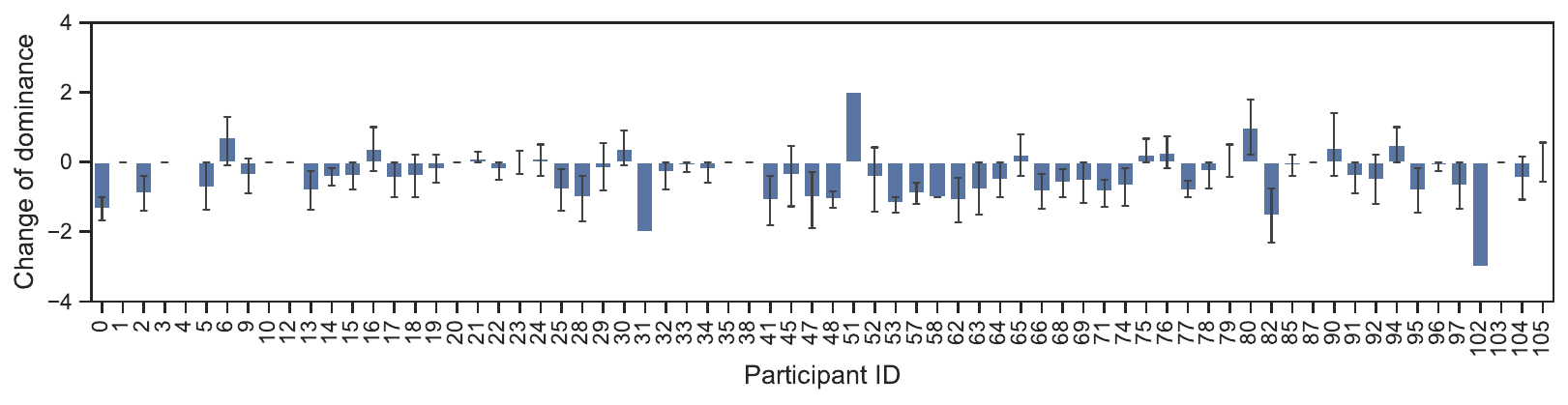}}
    \caption{Mean emotion shifts after homework involvement. The error bars represent the 95\% confidence interval.} 
    \label{fig: daily_survey_emotion}
\end{figure}

For the individual-level analysis, we calculated the mean emotional changes for each parent by subtracting their pre-homework emotional values from their post-homework values, as shown in Figure \ref{fig: daily_survey_emotion}. The average emotion shifts were: valence (mean = -0.467, std = 0.619), arousal (mean = 0.535, std = 0.575), and dominance (mean = -0.373, std = 0.686). We observed substantial variability across participants. While most parents (e.g., P0, P58, P82) experienced a decrease in pleasure, an increase in arousal, and a decrease in dominance, a few parents, such as P90, showed more stable emotional states, possibly demonstrating resilience. However, we acknowledge that while the emotional shift values are calculated using the group means for each participant, the varying number of responses contributed by different participants (see Appendix Figure \ref{fig:valence_count}) may influence these individual differences.

%For the individual-level analysis, we calculated the mean emotional change for each parent by subtracting their pre-homework emotional values from their post-homework values, as shown in Figure \ref{fig: daily_survey_emotion}. The average emotional shifts were: valence (mean = -0.47, SD = 0.62), arousal (mean = 0.53, SD = 0.58), and dominance (mean = -0.37, SD = 0.69). We observed substantial variability across participants. While most parents (e.g., P0, P58, P82) experienced a decrease in pleasure, an increase in arousal, and a decrease in dominance, a few parents (e.g., P90) showed more stable emotional states, possibly demonstrating resilience. However, we acknowledge that the number of responses contributed by different participants varied (see Appendix Figure \ref{fig:valence_count}), which may influence these individual differences. These findings highlight the diverse emotional responses of parents to homework involvement.

\subsection{Emotion Fluctuations During Homework Involvement} 
\subsubsection{Extracting Emotion Annotations}

%To gain a nuanced understanding of the emotional fluctuations of both parents and children during homework interactions, we employed the recently developed emotion fine-tuning large language model, EmoLLMs \cite{liu2024emollms}. This suite of models and annotation tools excels at comprehensive affective analysis, showing outstanding performance in emotion regression and classification tasks, particularly using the three dimensions—Pleasure, Arousal, and Dominance—from the EmoBank dataset \cite{buechel2022emobank}. Given these strengths, we utilized EmoLLMs to annotate our experiment's transcribed data.

To understand the emotional fluctuations during homework interactions, we employed the EmoLLMs model suite \cite{liu2024emollms}, which excels in affective analysis, particularly in emotion regression and classification tasks using the three dimensions (i.e., \textit{Pleasure}, \textit{Arousal}, and \textit{Dominance}) based on the EmoBank dataset \cite{buechel2022emobankstudyingimpactannotation}. Given its high performance, especially in \textit{Pleasure} analysis (Pearson Correlation Coefficient = 0.728), we used the EmoLLaMA-chat-7B model \cite{liu2024emollms} to annotate our transcribed data.

\begin{figure}
    \centering
    \includegraphics[width=0.99\linewidth]{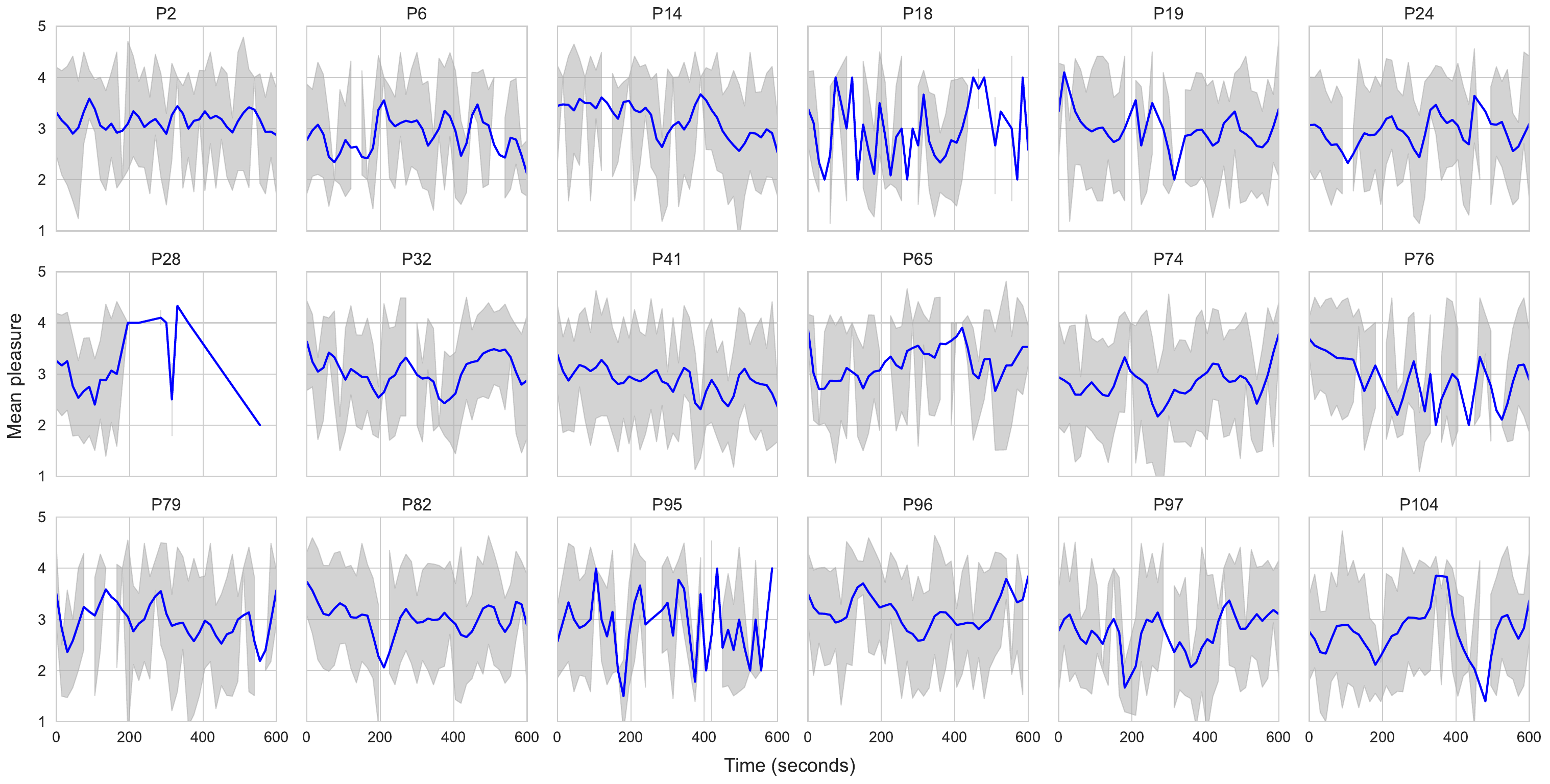}
    \caption{Average pleasure for the first 10 minutes of sessions. The dark grey band indicates the standard error.}
    \label{fig:emotion_all}
\end{figure}

We first filtered the transcriptions to remove sentences lacking semantic clarity, such as short or misrecognized sentences, which could distort the results. We then applied EmoLLMs to infer emotional dimensions on a sentence-by-sentence basis, focusing on the \textit{Pleasure} dimension. This is because pleasure directly reflects the emotional polarity (positive or negative) critical for understanding parent-child interactions during homework \cite{pekrun2002academic}. Although future analyses may include Arousal and Dominance, current model limitations make it practical to focus on pleasure alone for accurate analysis.

\subsubsection{Emotional Variation Analysis}

We collected emotional data from multiple homework sessions, totalling 40,356 \added{pleasure estimation} from 66 parents.
To capture the dynamics, we calculated the mean pleasure for each participant at 15-second intervals, a balance that retained key details without being overwhelmed by data limitations.  
To account for varied session lengths, we standardised our analysis by focusing on the first 10 minutes of each session, which aligns with similar studies \cite{tag2022emotion}. This allowed us to analyse emotional fluctuations across participants consistently. Although some sessions lasted longer than 10 minutes and others shorter, we used all available data to calculate the mean pleasure for the first 10 minutes of each session.

Figure \ref{fig:emotion_all} shows the average pleasure levels during the first 10 minutes, with LOESS smoothing (frac = 0.1) applied to capture the overall trend. The dark grey bands represent the standard error and are based on smoothed mean values rather than raw data to highlight clearer patterns amid the variability. Due to space constraints, we present data for 18 participants with the most \added{pleasure estimation}. We found that some parents exhibit an early decline in pleasure within the first 2-3 minutes (e.g., P32, P76, P82), while others show fluctuating pleasure levels throughout the first 10 minutes (e.g., P18, P95). This provides insights into the evolving emotional engagement of parents during the early phases of homework involvement.

\iffalse
\begin{figure}
    \centering
    \includegraphics[width=0.99\linewidth]{figure/dis_all.pdf}
    \caption{Average pleasure for sessions, with lengths normalised to [0,1]. The dark grey band indicates the standard error.}
    \label{fig:emotion_all_normalised}
\end{figure}
\fi

\section{Understanding Parents' Behavioural Dynamics During Homework Involvement}

%Parental behavioural dynamics play a crucial role in shaping the quality and effectiveness of homework assistance. Understanding these dynamics not only allows us to evaluate how parental actions influence a child’s academic development but also helps in identifying patterns that may lead to tension or conflict. By studying these dynamics, we gain insights into how different types of parental involvement can either support or hinder the child’s homework experience \cite{eccles1993parent,eccles2013family}.

In this section, we explore the parental behaviours and parent-child conflicts that arise during homework sessions. Using GPT-4o, we systematically analyse and categorise these behaviours and conflicts from transcribed conversations with the assistance of educational experts. We then validate the extracted behaviours and conflicts through comparisons with human annotations. Additionally, we examine the distribution of these behaviours and conflicts both across the overall population and on a per-user basis.

\subsection{Parental Behaviours and Parent-Child Conflicts}

%To establish a coding manual for parental behaviours and conflicts observed during homework assistance, we begin by defining these two key elements. \textbf{Parental behaviours} encompass the actions or responses of parents as they assist their children with homework. These behaviours can be positive, neutral, or negative depending on their nature and effect on the child. \textbf{Parent-child conflicts} refer to disagreements or tension that arise during homework sessions, which can be triggered by a range of factors, such as differing perspectives on learning methods or time management.

Although prior research has extensively explored parental behaviours and conflicts in general family and educational settings \cite{moe2018brief,solomon2002helping,nnamani2020impact}, to the best of our knowledge, there are no established definitions or a comprehensive codebook specifically addressing parental behaviours and parent-child conflicts during homework involvement. Therefore, we adopted a bottom-up coding process to inductively derive patterns from the raw data. This process is divided into three main steps: open coding, axial coding, and selective coding, ultimately resulting in the creation of a codebook for understanding parental behaviours and parent-child conflicts during homework sessions \cite{strauss1994grounded,glaser2017discovery,charmaz2006constructing}.

%Therefore, we begin by grounding our definitions in related literature on parenting and education but adapt them to the specific context of homework involvement. In cases where existing definitions are inadequate, we propose new ones based on our analysis of the interactions observed in this study.
 %coding manual

In this context, parental behaviour refers to the specific actions, responses, or strategies parents employ while assisting their children with homework. These behaviours include verbal and non-verbal interactions, guidance, feedback, or any form of involvement that influences the child's approach to homework. Parental behaviours may support, hinder, or remain neutral in the homework process, and they reflect the dynamic interaction between parent and child during educational activities \cite{cunha2015parents,eccles2013family}.
Similarly, parent-child conflicts are defined as moments of disagreement, tension, or friction that arise during homework involvement. These conflicts may stem from misunderstandings, differences in expectations, frustration, or emotional reactions from either the parent or the child, ranging from minor verbal disagreements to more disruptive disputes \cite{grolnick2009issues,benckwitz2023reciprocal,hanh2023current}.

%Similarly, parent-child conflicts during homework are defined as moments of disagreement, tension, or friction. These conflicts may arise from misunderstandings, differing expectations, frustration, or emotional reactions from either the parent or child, ranging from minor disagreements to more disruptive disputes.

%We randomly selected 50 transcripts from a larger dataset of 602 valid conversations for initial analysis. Using GPT-4o, we conducted two open coding tasks: one to identify parental behaviours and another to capture specific parent-child conflicts. This stage of open coding generated a total of 932 conflict scenarios and 2,161 instances of parental behaviour, resulting in 330 conflict codes and 950 behaviour codes. These codes formed the basis for further analysis and refinement.

\begin{table}
\centering
\scriptsize
\caption{Code definitions and examples of positive, neutral, and negative behaviours.}
\label{tab:behaviours} 
\begin{tabular}{p{0.12\textwidth} p{0.40\textwidth} p{0.47\textwidth}}
\toprule
\textbf{Code Name} & \textbf{Code Definition} & \textbf{Example} \\ \midrule
\multicolumn{3}{c}{\textbf{\textit{Positive Behaviours}}}   \\\midrule
\textit{Encouragement (ENC)} & Parents provide verbal or behavioural support to encourage the child's effort and progress, boosting their confidence and motivation to overcome challenges. & \textit{"You've worked really hard, keep it up! I believe in you!"} \newline \textit{"Don't worry, let's take it step by step, you'll definitely get it."} \\ \hline
\textit{Labelled Praise (LP)} & Parents specifically highlight the child's particular action or achievement and offer praise, helping the child recognize their specific progress and strengths. & \textit{"You did great on this addition problem, no mistakes at all!"} \newline \textit{"Your handwriting is especially neat this time, keep it up!"} \\ \hline
\textit{Unlabelled \newline Praise (UP)} & Parents give general praise to the child without pointing out specific actions or achievements. & \textit{"You're amazing, keep going!"} \newline \textit{"Wow, that's awesome!"} \\ \hline
\textit{Guided \newline Inquiry (GI)} & Parents ask questions or provide clues to guide the child toward independent thinking and problem-solving. & \textit{"Where do you think this letter should go?"} \newline \textit{"What methods could we use to solve this problem? Think about a few ways."} \\ \hline
\textit{Setting Rules (SR)} & Parents set clear rules or requirements for completing homework, helping the child establish good study habits and time management skills. & \textit{"You need to finish your Chinese homework before watching cartoons."} \newline \textit{"All homework needs to be done before dinner if you want to go out and play."} \\ \hline
\textit{Sensitive \newline Response (SRS)} & Parents respond to the child's emotions, needs, and behaviours in a timely, appropriate, and caring manner. & \textit{"I can see you're a bit tired now, how about we take a break and continue later?"} \newline \textit{"Do you find this question difficult? Don't worry, let's take another look together."} \\ \midrule
\multicolumn{3}{c}{\textbf{\textit{Neutral Behaviours}}}   \\\midrule

\textit{Direct \newline Instruction (DI)} & Parents tell the child how to complete a task or solve a problem without using guided or inquiry-based methods. & \textit{"For this problem, you should do it like this: add 4 to 6 to get 10."} \newline \textit{"Just copy this answer down, don't overthink it."} \\ \hline
\textit{Information \newline Teaching (IT)} & Parents teach new knowledge or skills by explaining concepts, reading texts, or offering detailed instructions. & \textit{"The character 'tree' is written with a wood radical on the left and 'inch' on the right, let's write it together."} \newline \textit{"You need to memorize multiplication tables like this: two times two equals four, two times three equals six. Let's start with those."} \\ \hline
\textit{Error Correction (EC)} & Parents point out mistakes in the child's homework and guide them to correct or revise their work. & \textit{"You missed the 'wood' radical here, write it again."} \newline \textit{"The addition is wrong here, let's calculate it again. Remember to line up the numbers correctly."} \\ \hline
\textit{Monitoring (MON)} & Parents regularly check the child's homework progress or completion to ensure they stay on track. & \textit{"How many pages have you written? Let me check for mistakes."} \newline \textit{"Let me look over your pinyin homework today to see if everything is correct."} \\ \hline
\textit{Direct Command (DC)} & Parents use clear and direct language to request or command the child to perform a specific action or task. & \textit{"Go do your math homework right now, no more delays!"} \newline \textit{"Stop playing with your toys and go finish your pinyin practice."} \\ \hline
\textit{Indirect Command (IC)} & Parents indirectly convey their requests, often through suggestions or hints, rather than giving direct orders. & \textit{"Have you finished your homework? Maybe it's time to get it done."} \newline \textit{"How about we finish homework first and then go play? That way you won't have to worry about running out of time later."} \\ \midrule

\multicolumn{3}{c}{\textbf{\textit{Negative Behaviours}}}   \\\midrule
\textit{Criticism and \newline Blame (CB)} & Parents express negative evaluations of the child's mistakes or behaviours, often directly blaming the child. & \textit{"How could you mess up such a simple word?"} \newline \textit{"I've told you this a thousand times, why haven't you remembered it yet?"} \\ \hline
\textit{Forcing and \newline Threatening (FT)} & Parents use pressure or threats of consequences to force the child to comply with their demands. & \textit{"If you don't do your homework, you won't be allowed to play with your toys today!"} \newline \textit{"If you don't finish, I'll take away your toys!"} \\ \hline
\textit{Neglect and \newline Indifference (NI)} & Parents show a lack of attention or emotional response to the child's needs or feelings. & \textit{\textbf{Child}: "Mom, I don't understand this question, can you help me?"} \newline \textit{\textbf{Parent}: (no response, continues using phone)} \\ \hline
\textit{Belittling and \newline Doubting (BD)} & Parents belittle the child's abilities or question their performance, undermining the child's confidence and motivation. & \textit{"How could you be so stupid? You can't even solve simple addition."} \newline \textit{"With grades like these, you'll never get into a good school."} \\ \hline
\textit{Frustration and \newline Disappointment (FD)} & Parents express frustration or disappointment in the child's performance when it fails to meet their expectations. & \textit{"I can't believe you did so poorly on this test, I'm really disappointed."} \newline \textit{"I thought you'd do better, but I guess I was wrong."} \\ \hline
\textit{Impatience and \newline Irritation (II)} & Parents exhibit impatience or irritation when the child's performance falls short of expectations. & \textit{"Why are you so slow? I've been waiting forever!"} \newline \textit{"Why isn't this finished yet? You always take so long!"} \\ \bottomrule
\end{tabular}
\end{table}

\subsubsection{Coding and Categorisation}

To conduct this study, we randomly sampled 50 transcripts from 602 valid dialogues. Using GPT-4o, we performed two open coding tasks: one to identify parental behaviours and another to capture specific parent-child conflicts. Through this process, we identified 932 conflict scenarios and 2,161 instances of parental behaviour, which initially resulted in 330 conflict codes and 950 behaviour codes. These preliminary codes provided a rich foundation for further analysis.

In the \textit{axial coding} phase, we refined the initial codes by removing those that did not reflect genuine conflicts and merging similar codes to reduce redundancy. This refinement process resulted in \textit{166 conflict codes} and \textit{606 behaviour codes}. Using GPT-4o, we then categorised these codes based on content similarity and emerging patterns, grouping the conflict codes into \textit{12 categories} and the behaviour codes into \textit{34 categories}. These categories represent typical conflict and behavioural patterns, offering a structured framework for analysing parent-child interactions during homework.

In the \textit{selective coding} phase, a human education expert reviewed ten parent-child dialogue samples and integrated the results from the axial coding stage. The expert focused on identifying core conflict types while considering the educational and cultural context of Chinese parents. Redundant or ambiguous codes were consolidated, reducing the \textit{12 conflict categories} to \textit{8 core conflict types} and the \textit{34 behaviour categories} to \textit{18 key behaviour categories}. Each conflict type and behaviour category was clearly defined, with representative dialogue examples chosen to illustrate their practical use.

Following this, three educational experts were invited to review the coding manual, evaluating its accuracy, relevance, and applicability. Based on their feedback, further refinements were made to ensure clarity and consistency in the code definitions and guidelines. After multiple revisions, the coding manual was finalised as a validated tool for future qualitative research, providing clear guidance for analysing parental behaviours and parent-child conflict during homework involvement.

%After developing the initial version of the coding manual, we sought the evaluation of \textit{three educational experts}, who reviewed the coding scheme in depth. These experts assessed the accuracy, relevance, and applicability of the codes within the educational context, providing constructive feedback for improvement. Based on their recommendations, we made further refinements to the code definitions and coding guidelines, optimizing the manual for clarity and consistency. Through multiple rounds of revision and expert input, we finalized the coding manual, which now serves as a validated tool for future qualitative research. This manual offers clear guidance for the coding process, ensuring consistency across different researchers while providing a robust foundation for analyzing the underlying causes and patterns of parent-child conflict during homework assistance.

%We applied GPT-4o to the entire dataset to automate the coding of parental behaviours and conflicts. In the case of \textbf{parental behaviour coding}, GPT-4o segmented each dialogue into distinct behaviour units, each mapped to a corresponding code based on predefined categories. Behaviours were classified as positive, neutral, or negative, with descriptions provided for each identified instance. Similarly, for \textbf{parent-child conflict coding}, GPT-4o identified and labelled conflict scenarios, categorizing them by the conflict’s trigger, development, intensity, and the reactions of both parent and child.
\begin{table}
\centering
\footnotesize
\caption{Conversations of parent-child conflicts in homework involvement, as synthesised by ChatGPT. Due to the lack of a comprehensive conflict coding manual specifically tailored for homework involvement scenarios, we conducted extensive literature reviews, interviewed educational experts, and utilised GPT to analyse and summarise common types of Parent-Child Conflicts in these contexts. After validation and refinement by educational experts, we developed specific definitions and examples for each conflict type. We then employed GPT to code our conversation transcripts accordingly.}
\label{tab:conflict}
\begin{tabular}{p{0.10\textwidth} p{0.53\textwidth} p{0.33\textwidth}}
\toprule
\textbf{Code Name} & \textbf{Code Definition} & \textbf{Example} \\ 
\midrule
\textit{Expectation Conflict (EC)} & This conflict arises when parents have high expectations for their child's performance, progress, or future, but the child’s actual abilities, goals, or interests do not align with these expectations. Parents may also compare their child to others, intensifying the conflict. & \textit{\textbf{Parent}: “You should be like your classmate and get full marks! How could you get such an easy question wrong?”} \newline \textit{\textbf{Child}: “I’ve done my best. Why do you always think I’m worse than others?”} \\

\midrule
\textit{Communication Conflict (CC)} & This conflict arises when parents and children have different communication styles during homework sessions, leading to misunderstandings and emotional tension. Parents may criticize, question, or belittle the child, making the child feel misunderstood or oppressed, thus escalating the communication barrier. & \textit{\textbf{Parent}: “What’s wrong with you? I’ve explained this so many times and you still don’t get it!”} \newline \textit{\textbf{Child}: “I just don’t want to listen to you anymore. You always yell at me!”} \\

\midrule
\textit{Learning Method \newline Conflict (LMC)} & This conflict occurs when parents and children disagree on how to approach and complete homework. Parents may feel the child’s method is inefficient and try to impose their own approach, while the child insists on using their own method and resists parental intervention. & \textit{\textbf{Parent}: “You shouldn’t study like this. Finish all the questions first, then check your answers!”} \newline \textit{\textbf{Child}: “I’m used to doing it my way. Why should I follow what you say?”} \\

\midrule
\textit{Rule Conflict (RC)} & This conflict occurs when parents set strict rules for learning, and the child seeks more autonomy. Parents may try to control the pace or structure of the child’s study sessions, while the child resists these restrictions and pushes for greater flexibility and freedom. & \textit{\textbf{Parent}: “You must start your homework right after dinner. No more delays!”} \newline \textit{\textbf{Child}: “I want to play for a little longer. You’re always controlling everything!”} \\

\midrule
\textit{Time \newline Management Conflict (TMC)} & This conflict arises from disagreements about how time and energy should be allocated for studying. Parents may want the child to follow a fixed study schedule, while the child may prefer a different routine, leading to conflict. & \textit{\textbf{Parent}: “You always leave your homework until so late at night. You’re so inefficient!”} \newline \textit{\textbf{Child}: “I prefer studying later. I just can’t focus in the morning!”} \\

\midrule
\textit{Knowledge Conflict (KC)} & This conflict occurs when there is a mismatch in knowledge levels or understanding between parents and children. Parents may have already mastered certain knowledge and find it difficult to empathize with the child’s struggles, or they may explain concepts from a perspective the child cannot yet grasp. Additionally, parents might be unfamiliar with certain subjects, leading the child to question their guidance. & \textit{\textbf{Parent}: “This problem is so simple. How can you still not understand it?”} \newline \textit{\textbf{Child}: “You don’t understand what I’m struggling with! My teacher explained it differently from you.”} \\

\midrule
\textit{Focus Conflict (FC)} & This conflict arises when parents are dissatisfied with the child’s attention or focus during study time, believing the child is distracted or not concentrating sufficiently. Parents may attempt to intervene or remind the child to focus, while the child may feel pressured or overwhelmed by the interference, leading to emotional conflict. & \textit{\textbf{Parent}: “What are you daydreaming about? Focus on your homework!”} \newline \textit{\textbf{Child}: “I wasn’t distracted, I was just thinking about how to solve the problem.”} \\

\bottomrule
\end{tabular}
\end{table}
\subsubsection{Automated Coding Using GPT-4o}
We applied GPT-4o to the entire dataset to automate the coding of parental behaviours and conflicts. Regarding the parental behaviour coding, we utilise GPT-4o to analyse specific actions of the parent during the homework involvement process. Each segment identified distinct behaviours, segmenting the dialogue into behaviour units (denoted as \texttt{behaviour\_id}), where each behaviour could consist of one or more sentences. It then provided a brief description for each behaviour and mapped it to a relevant \textit{code} based on predefined categories. These behaviours were classified as \textit{positive}, \textit{neutral}, or \textit{negative}, ensuring accurate classification of each parental action. The specific coding manual is presented in Table \ref{tab:behaviours}, and the coding guidelines are detailed in Table \ref{appen:tab:behaviour}.

For the parent-child conflict coding, GPT-4o was prompted to identify conflict scenarios from the transcribed dialogues. Each conflict was described according to the following dimensions: \textit{trigger of the conflict}  (what initiated it), \textit{development} (how it unfolded), \textit{parent’s behaviour} (verbal or non-verbal reactions), \textit{child’s response}, \textit{type of conflict}, and \textit{its intensity} (categorised as high, medium, or low). GPT-4o assigned a short label, or code, to each dialogue segment representing a conflict. The specific coding manual for these conflicts is presented in Table \ref{tab:conflict}, and the coding guidelines are provided in Table \ref{appen:tab:conflict}. \added{The representative examples in Tables \ref{tab:behaviours} and \ref{tab:conflict} were automatically selected by GPT-4o within each code category to illustrate the typical manifestations of each behaviour or conflict type.}

\subsection{Evaluation of Coding Consistency}

%To assess the consistency of GPT-4o's coding, we conducted an evaluation experiment comparing the AI-generated codes with those created by human experts. We randomly selected 200 instances from each coding task for detailed analysis and asked four human experts to code the same instances independently. Their coding results were compared with GPT-4o's outputs using \textit{Cohen's Kappa Coefficient} to measure the agreement. A \textit{Consensus Coding} was established via majority voting among experts, serving as the gold standard for comparison with GPT-4o. The experiment evaluated consistency across three dimensions: between human experts, between GPT-4o and individual experts, and between GPT-4o and the expert consensus.

To evaluate the consistency of GPT-4o's coding, we conducted an experiment comparing its outputs with those of human experts. From each coding task, 200 instances were randomly selected. Four human experts independently coded these instances, and their results were compared with GPT-4o's using \textit{Cohen's Kappa Coefficient}. A \textit{Consensus Coding} was derived through majority voting among the experts and served as the gold standard for comparison. We evaluated consistency across three dimensions: between human experts, between GPT-4o and individual experts, and between GPT-4o and the expert consensus.

\begin{table}
    \footnotesize
    \centering
    \caption{Cohen's Kappa consistency analysis between GPT-4o and human experts. This table shows the inter-rater agreement (Cohen’s Kappa) between human experts (Expert 1 to Expert 4) and GPT-4o. Each value represents the level of agreement between expert pairs or between GPT-4o and individual experts. The final column provides Kappa values comparing GPT-4o to the consensus coding derived from majority voting among experts. In case of a tie, an additional education expert was consulted for arbitration. This consensus coding serves as the benchmark for consistency. Higher Kappa values indicate stronger agreement.}
    \label{tab:consistency_analysis} 
    \begin{tabular}{llccccc}
        \toprule
        \multirow{2}{*}{\textbf{Coding Task}} & \multirow{2}{*}{\textbf{Expert}} & \multicolumn{4}{c}{\textbf{Cohen's Kappa Value}} & \multirow{2}{*}{\textbf{Kappa Value w/ Consensus}} \\
        \cmidrule(lr){3-6}
        &  & Expert 1 & Expert 2 & Expert 3 & Expert 4 & \\
        \midrule
        \multirow{6}{*}{\textit{Conflict Coding}} 
        & Expert 1 & -- & 0.330 & 0.348 & 0.409 & \textbf{0.519} \\
        & Expert 2 & 0.330 & -- & 0.405 & 0.392 & \textbf{0.510} \\
        & Expert 3 & 0.348 & 0.405 & -- & 0.738 & \textbf{0.768} \\
        & Expert 4 & 0.409 & 0.392 & 0.738 & -- & \textbf{0.804} \\
        & \textbf{GPT-4o} & \textbf{0.410} & \textbf{0.444} & \textbf{0.458} & \textbf{0.500} & \textbf{0.517} \\
        \midrule
        \multirow{6}{*}{\textit{Behaviour Coding}} 
        & Expert 1 & -- & 0.473 & 0.457 & 0.467 & \textbf{0.587} \\
        & Expert 2 & 0.473 & -- & 0.524 & 0.627 & \textbf{0.697} \\
        & Expert 3 & 0.457 & 0.524 & -- & 0.802 & \textbf{0.797} \\
        & Expert 4 & 0.467 & 0.627 & 0.802 & -- & \textbf{0.852} \\
        & \textbf{GPT-4o} & \textbf{0.574} & \textbf{0.708} & \textbf{0.562} & \textbf{0.653} & \textbf{0.724} \\
        \bottomrule
    \end{tabular}
\end{table}

As shown in Table \ref{tab:consistency_analysis}, GPT-4o demonstrated \textit{moderate to substantial agreement} with individual experts in the behaviour coding task, with Kappa values ranging from \textbf{0.562} (Expert 3) to \textbf{0.708} (Expert 2). According to Landis and Koch \cite{landis1977measurement}, these values indicate \textit{substantial agreement} with Expert 2 and \textit{moderate agreement} with Expert 3. Agreement with the expert consensus was even higher, at \textbf{0.724}, indicating \textit{substantial agreement}. In the conflict coding task, Kappa values with individual experts ranged from \textbf{0.410} (Expert 1) to \textbf{0.500} (Expert 4), and the value against consensus was \textbf{0.517}, all within the \textit{moderate agreement} range. Overall, GPT-4o’s consistency with the expert consensus was comparable to that of individual human annotators across both tasks.

%The moderate agreement in conflict coding is reasonable, given that dialogue act annotation and sentiment analysis tasks often involve subjective interpretations, as noted in previous research. Studies have highlighted that these tasks are inherently complex and prone to ambiguity, especially when dealing with conflicting or emotionally nuanced data. For instance, Stolcke et al. \cite{stolcke2000dialogue} discuss the challenges in annotating dialogue acts, pointing out that annotators may struggle with the subtlety and context-dependency of dialogues, leading to moderate or low Kappa values. 
The moderate agreement in conflict coding is reasonable given the subjective nature of the task. Prior work, such as Stolcke et al. \cite{stolcke2000dialogue}, highlights the inherent ambiguity in dialogue act annotation, where subtle shifts in tone and context can lead to annotator disagreement. 
Similarly, \added{Latif et al. \cite{latif2023can}}, and Litman et al. \cite{litman2003recognizing} argue that disagreement among annotators is common, even with proper training, especially in tasks involving emotion or sentiment, where subtle cues may be interpreted differently. These studies suggest that a Kappa range of 0.41 to 0.60 is typical in such contexts, underlining the subjective nature of the task.

%The moderate agreement in conflict coding is reasonable given the subjective nature of the task. Prior work, such as Stolcke et al. \cite{stolcke2000dialogue}, highlights the inherent ambiguity in dialogue act annotation, where subtle shifts in tone and context can lead to annotator disagreement. Similarly, Latif et al. \cite{latif2023can} and Litman et al. \cite{litman2003recognizing} note that even with training, sentiment and emotion-related annotations often yield only moderate Kappa values due to their nuanced nature. In this context, a Kappa range between 0.41 and 0.60 is typical and reflects the complexity of the coding task rather than poor reliability.

%To better understand where GPT-4o aligned well with expert coding and where discrepancies occurred, we employed confusion matrices to analyse the specific categories of errors and misclassifications. These matrices, detailed in Appendix \ref{sec:Confusion}, provide patterns of agreement and confusion across different coding categories, highlighting strengths and areas needing improvement.

To better understand where GPT-4o aligned with or diverged from expert annotations, we analysed confusion matrices (Appendix \ref{sec:Confusion}). These matrices highlight patterns of agreement and common misclassifications across categories, offering insights into the model’s strengths and areas for refinement.
%\added{In addition, we conducted a \textit{Chi-Squared} test to evaluate whether the observed distribution of GPT-4o's coding significantly deviates from what would be expected by chance. Specifically, the test examines whether the distribution of entries in the confusion matrix (i.e., the joint frequencies of GPT-4o's labels and expert annotations) differs from the distribution we would expect if GPT-4o’s predictions were independent of the expert labels. In other words, the null hypothesis assumes that GPT-4o's coding decisions are unrelated to the expert coding.}\added{For behaviour coding, the Chi-Squared statistic was $\chi^2 = \text{1756.134}$ with a p-value of $1.19 \times 10^{-221}$, indicating an extremely significant deviation from independence. Similarly, for conflict coding, $\chi^2 = \text{333.131}$ with a p-value of $8.39 \times 10^{-50}$. These results strongly reject the null hypothesis, confirming that GPT-4o's predictions are not random and that its coding aligns with expert annotations to a statistically significant degree, thereby further validating the model’s reliability.}
We also conducted a \textit{Chi-Squared} test to assess whether GPT-4o’s coding deviated significantly from random chance. The test examined whether the distribution of predicted labels was independent of expert annotations. For behaviour coding, the result was $\chi^2 = 1756.134$, with a p-value of $1.19 \times 10^{-221}$. For conflict coding, $\chi^2 = 333.131$ with a p-value of $8.39 \times 10^{-50}$. Both results reject the null hypothesis with extremely high significance, confirming that GPT-4o’s predictions are strongly aligned with expert coding and not random.

In sum, the evaluation experiments show that LLM-based coding achieves \textit{substantial agreement} in behaviour coding and \textit{moderate agreement} in conflict coding by human expert annotators. While it  aligns closely with the expert consensus, some discrepancies remain, particularly with Expert 1. Further refinement in conflict coding could improve its performance in more nuanced or complex interactions.

\subsection{Analysis of Behavioural Dynamics}

\subsubsection{Overall Trend}

\begin{figure}
    \centering
    \begin{minipage}[t]{0.525\textwidth}
        \includegraphics[width=1\textwidth]{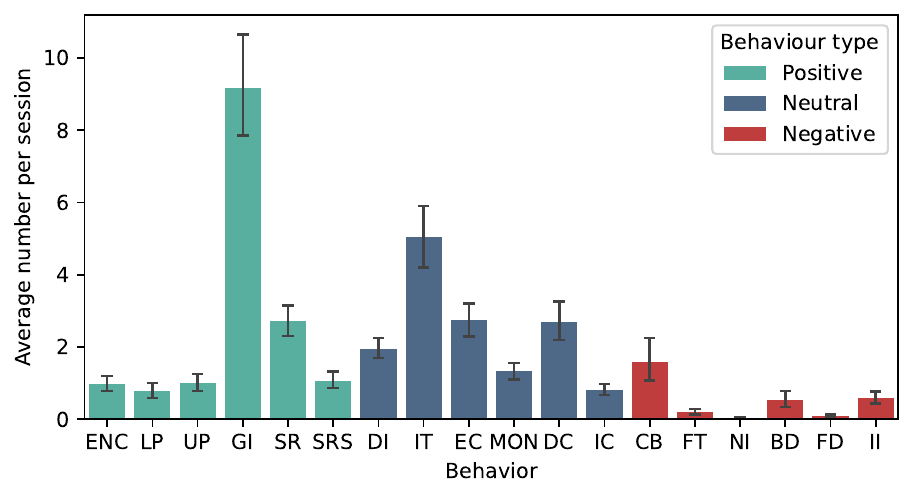}
        \caption{Average number of positive, neutral, and negative behaviours for all participants, with positive behaviours (e.g., Encouragement
(ENC)), neutral behaviours (e.g., Direct
Instruction (DI)), and negative behaviours (e.g., Criticism and
Blame (CB)) defined in Table \ref{tab:behaviours}. Error bar indicates 0.95 confidence level.}
        \label{fig:dis_behaviour}
    \end{minipage}
    \hspace{0.2cm}
    \begin{minipage}[t]{0.445\textwidth}
    \centering
        \includegraphics[width=1\textwidth]{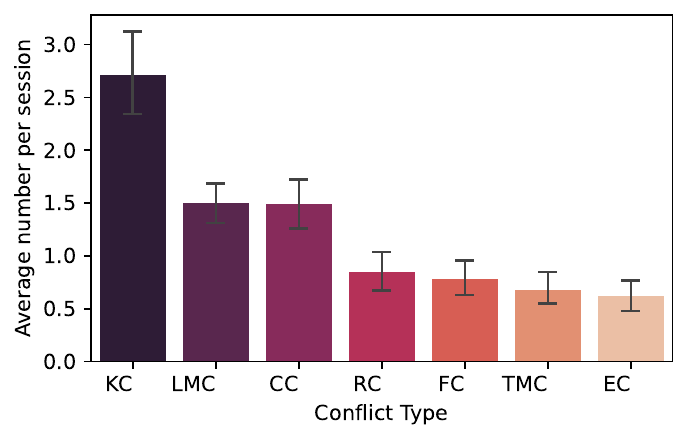}
        \caption{Average numbers of each conflict type per user, including types of conflicts defined in Table \ref{tab:conflict} (e.g., Knowledge Conflict (KC), Communication Conflict (CC), and Focus Conflict
(FC)). Error bar indicates 0.95 confidence level.}
        \label{fig:dis_conflict}
    \end{minipage}
    
\end{figure}

For each homework involvement session, we extracted parental behaviours and parent-child conflicts. Figure \ref{fig:dis_behaviour}
illustrates the average number of positive, neutral, and negative behaviours per session across all participants. Among these categories, positive behaviours were the most prevalent, accounting for 47.21\% of total behaviours, while negative behaviours were the least frequent at 9.07\%. The most commonly exhibited behaviour was \textit{Guided Inquiry}, with parents engaging in this behaviour an average of 9.16 times per session. This indicates that parents often ask questions or provide clues to guide their children towards independent thinking and problem-solving. In contrast, negative behaviours occurred less frequently, with parents displaying an average of 3.02 negative behaviours per session. Among these, \textit{Criticism and Blame} was the most common negative behaviour.

Despite the predominance of positive and neutral behaviours, the occurrence of parent-child conflicts was frequent, with an average of 8.63 per session. Figure \ref{fig:dis_conflict} highlights \textit{Knowledge Conflict} as the most common conflict (2.71 times per session), possibly due to the gap between parents’ higher education levels and their children’s earlier stages of learning. This aligns with the \textit{Curse of Knowledge} bias \cite{birch2007curse}, where experts struggle to understand novice perspectives. The second most common conflict type was \textit{Learning Method Conflict}, averaging 1.50 times per session, likely due to the disagreements between parents and children over the approach or strategies used for learning and completing homework tasks. 

\subsubsection{Individual Analysis}
\begin{figure}
    \centering
    \includegraphics[width=.95\linewidth]{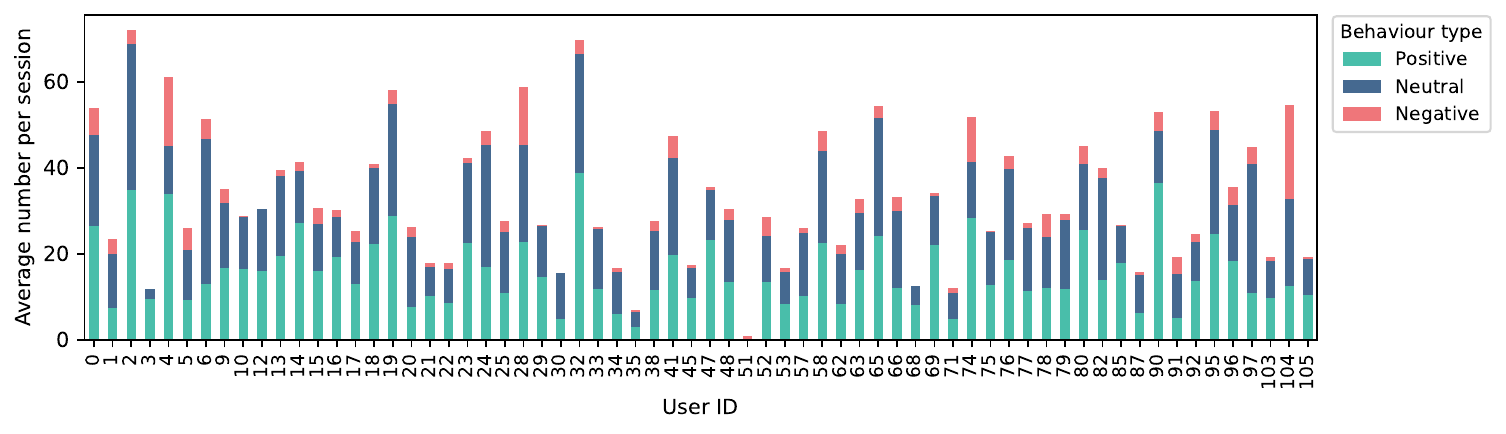}
    \caption{Average frequency of positive, neutral, and negative behaviours per homework session for each parent.}
    \label{fig:per_behaviour_dis}
\end{figure}

To further understand negative parental behaviours, we analysed their distribution across individual participants. Figure \ref{fig:per_behaviour_dis} presents the average number of behaviours per participant. While some participants exhibited a very small proportion of negative behaviours, others, such as P4, P28, and P104, displayed a significantly higher percentage of negative behaviours compared to their neutral and positive behaviours. This suggests that these individuals may benefit from improving their behaviours during homework involvement.

\begin{figure}
    \centering
    \includegraphics[width=1\linewidth]{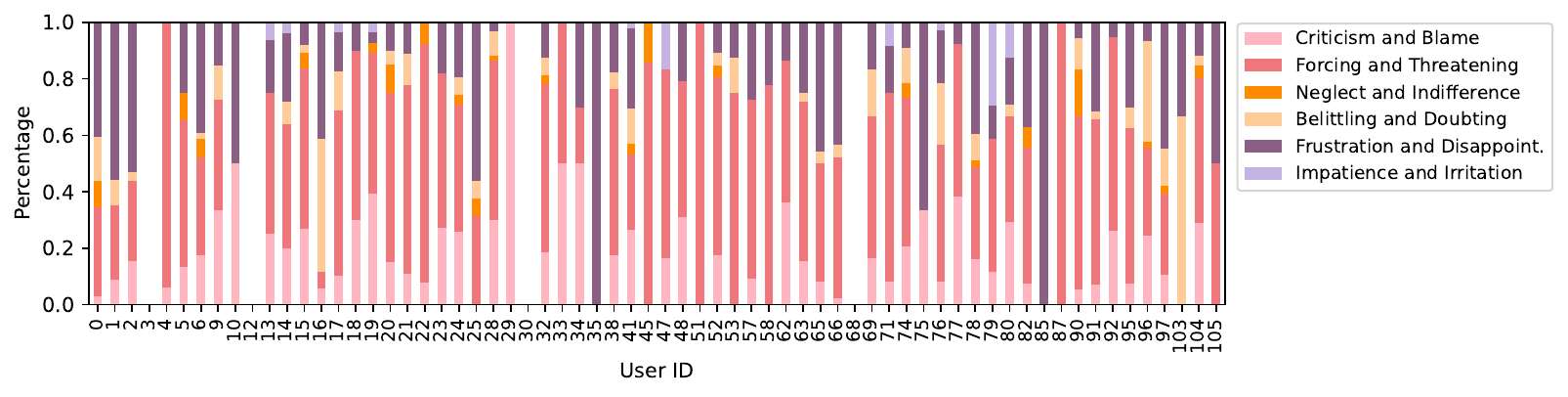}
    \caption{Proportion of various negative behaviours per homework session for each parent.}
    \label{fig:per_behaviour_stacked}
\end{figure}

We then examined the specific types of negative behaviours exhibited by each participant, as illustrated in Figure \ref{fig:per_behaviour_stacked}. We found that some participants demonstrated only limited types of negative behaviours. For example, P4 displayed just two types of negative behaviours, with \textit{Forcing and Threatening} accounting for 93.75\% of all negative behaviours. Similarly, P33 exhibited two types of negative behaviours, with \textit{Criticism and Blame} and \textit{Forcing and Threatening} each contributing 50\%. In contrast, participants such as P0, P6, P32, and P104 displayed a broader range of negative behaviours, suggesting a more varied pattern of interaction. These variations highlight the individual differences in how parents manage homework involvement and may point to areas for targeted behavioural improvement.
 
\begin{figure}
    \centering
    \includegraphics[width=.95\linewidth]{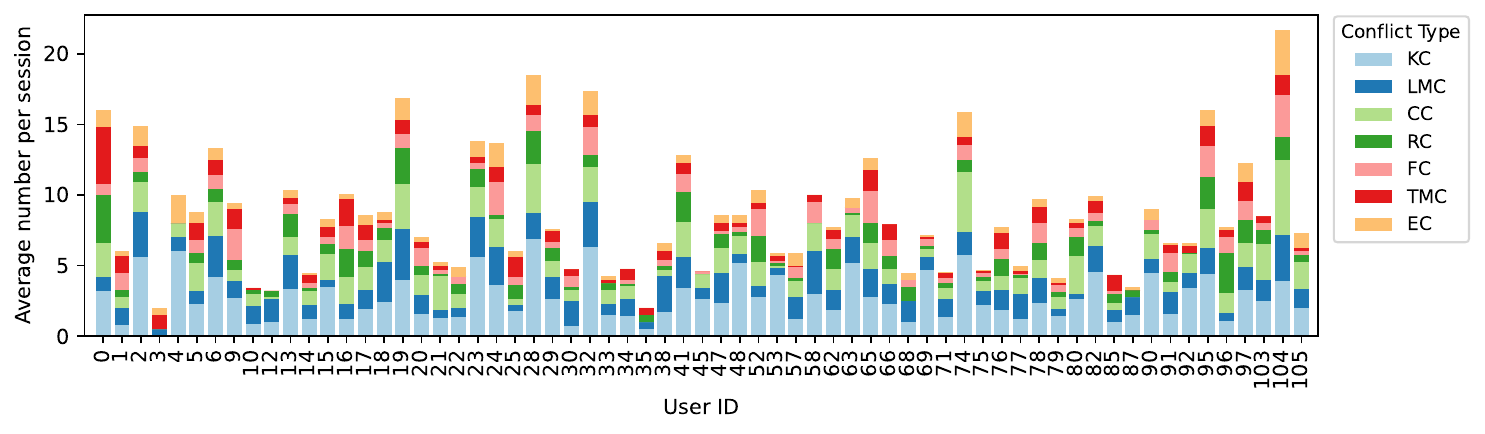}
    \caption{Average frequency of various conflict types per homework session for each parent.}
    \label{fig:dis_stacked_conflict}
\end{figure}

Finally, we investigated the average number of different conflict types per homework involvement session for each parent, as shown in Figure \ref{fig:dis_stacked_conflict}. Our findings reveal significant variability in the number of conflicts experienced by different families. For example, participants like P28, P32, and P104 averaged more than 15 conflicts per session, while others, such as P3 and P35, experienced fewer than 3 conflicts on average. Moreover, the variety of conflict types also differed among participants. Some, like P3, encountered only 3 types of conflict, while others such as P4, P35, and P87 experienced 4 types. In contrast, participants like P0, P16, and P28 encountered all types of conflicts. This suggests that the complexity of conflict during homework sessions can vary greatly among families. Although \textit{Knowledge Conflict} was the most common type of conflict overall, some participants experienced other types more frequently. For instance, P0 primarily dealt with \textit{Time Management Conflict}, while P96 had \textit{Rule Conflict} as their dominant issue. These findings highlight that not only the number of conflicts but also the types of conflicts vary widely among families.

\section{Relationships Between Emotions, Behaviours and Conflict in Chinese Families}

This section explores the intricate relationships among emotions, behaviours, and conflicts during homework involvement in Chinese families. We analysed data from 511 homework sessions across 65 families (excluding one due to insufficient data), computing the total occurrences of each behaviour and conflict type per session. Emotional shifts were assessed based on the change in parents’ self-reported pleasure, arousal, and dominance before and after the homework.

\begin{figure}
    \centering
    \includegraphics[width=1\linewidth]{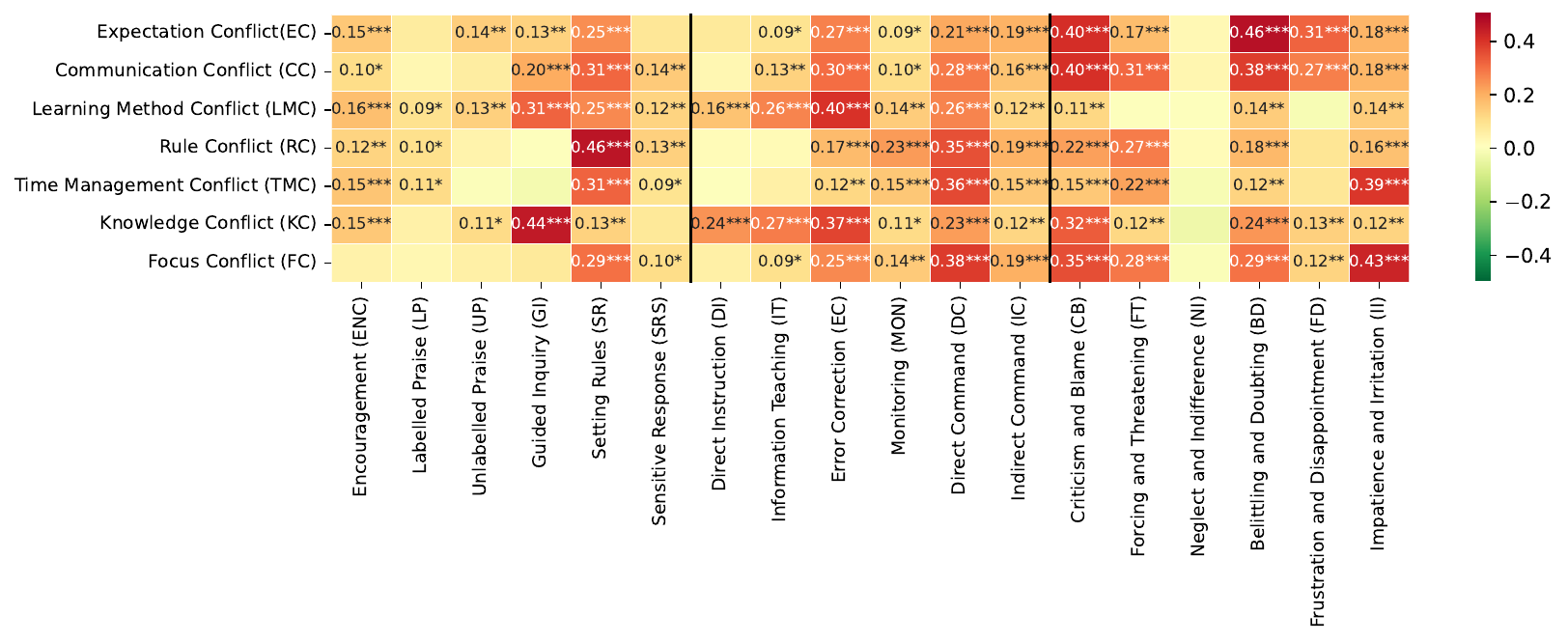}
    \caption{Correlation between parental behaviours and parent-child conflicts (* p<0.05, ** p<0.01, *** p<0.001).}
    \label{fig:heatmap:conflict_behaviours}
\end{figure}

Figure \ref{fig:heatmap:conflict_behaviours} displays the correlation matrix between parental behaviours and conflicts, with behaviours categorised as positive, neutral, or negative (delineated by bold lines). Surprisingly, significant positive correlations with conflict are found across all behaviour types, including positive behaviours, such as \textit{Labelled Praise} and \textit{Unlabelled Praise}. This counterintuitive finding aligns with family psychology research suggesting that well-intentioned parental involvement can sometimes increase tension during learning activities \cite{cunha2015parents}. For example, Cheung et al. \cite{cheung2016controlling} and Pomerantz et al. \cite{pomerantz2007whom} observed that controlling forms of support, even when positively intended, may undermine children’s autonomy. Similarly, praise can be perceived as evaluative rather than supportive, potentially triggering defensive responses from children \cite{ryan1982control}.

To further illustrate this, we conducted a case-level analysis for an example dialogue (see Table \ref{tab:cased}). This example, from Participant 0 on 11 April 2024, highlights how encouragement can become entangled with conditional approval and pressure. The exchange begins with positive reinforcement (\textit{"keep it up, great job"}) but evolves into implicit emotional contingencies. The child's comment (\textit{“But when you want to get angry, you still get angry at me”}) reveals an acute awareness of these shifting dynamics. 
As Pomerantz et al. \cite{pomerantz2007whom} point out, such "controlling" encouragement may shift focus from learning to performance, raising pressure rather than motivation. In the example, the parent's initial encouragement (\textit{"come on, one final push"}) gradually escalates into pressure, with expectations shifting from task completion to flawless performance (\textit{"it also has to be all correct"}). This reflects the \textit{Moving Goalposts} phenomenon described by Gurland and Grolnick \cite{gurland2005perceived}, where shifting expectations can undermine children’s sense of competence. 
Such dynamics help explain why even seemingly positive behaviours may correlate with conflict, supporting Nunez et al.’s \cite{nunez2015relationships} argument that the quality and context of parental involvement matter more than the presence of traditionally “positive” behaviours.

\begin{table}
\centering
\footnotesize
\caption{Case dialogue of parent-child conversation during homework involvement with behaviour coding (Participant 0, April 11, 2024). 
"P" denotes the parent, and "C" denotes the child. Only the parent's behaviours are coded in the Behaviour column.}
\label{tab:cased}
\begin{tabularx}{\textwidth}{>{\bfseries}l X l}
\toprule
\added{Speaker} & \added{Utterance} & \added{Behaviour} \\
\midrule
\added{P} & \added{Great job on the first page—well done. Keep it up, great job, you're doing really well. You're really holding yourself to a high standard today.} & \added{LP} \\
\added{C} & \added{Yeah, I'm really proud of myself.} & \added{--} \\
\added{P} & \added{You should be! It shows you're growing up and starting to expect more from yourself.} & \added{ENC} \\
\added{C} & \added{But when you want to get angry, you still get angry at me.} & \added{--} \\
\added{P} & \added{If you have high standards for yourself, I don't need to get angry. But if you have very low standards for yourself, what do you think I'll do?} & \added{GI} \\
\added{C} & \added{Get angry.} & \added{--} \\
\added{P} & \added{So what do you want from me? For example, if you deliberately misbehave in front of me, don't do your homework, make a mistake with everything you write, can't correct things even after being told 5 times, do you think I'll be happy or angry?} & \added{SR, CB} \\
\added{P} & \added{You're almost out of time, it's almost time to finish all your homework, and you haven't completed this one task yet. Sigh, I really don't want to get angry with you, but if I don't watch you...} & \added{MON, FD} \\
\added{C} & \added{Can't you just explain it to me in a simpler way?} & \added{--} \\
\added{P} & \added{Have you timed yourself? How long will it take? Why are you hitting your face? Hurry up, hurry up. Once you finish this, today's tasks will be complete. Come on, one final push, one burst of energy. Good—5 minutes will be enough.} & \added{MON, IC, ENC} \\
\added{C} & \added{I think I can do it.} & \added{--} \\
\added{P} & \added{Good, but it also has to be all correct, okay?} & \added{SR, ENC} \\
\bottomrule
\end{tabularx}
\end{table}

As shown in Figure \ref{fig:heatmap:conflict_behaviours}, among all behaviours, only \textit{Neglect and Indifference} showed no significant correlation with any conflict. This echoes Pomerantz et al.'s \cite{birch2007curse} findings that excessive involvement can hinder autonomy. This raises key questions: Can a certain level of detachment reduce conflict during homework?  How might parents strike a balance between involvement and harmony during homework sessions? As Hoover-Dempsey and Sandler \cite{hoover2005social} argued, the challenge for parents is not simply being involved, but carefully navigating how they engage. 
Different behaviours are also associated with specific conflict types. Behaviours such as \textit{Setting Rules}, \textit{Error Correction}, and \textit{Criticism and Blame} are linked to all conflict types, suggesting that even corrective or guiding intentions may inadvertently increase friction. On the other hand, behaviours like \textit{Labelled Praise}, \textit{Unlabelled Praise}, and \textit{Direct Instruction} show fewer conflict correlations, though not entirely conflict-free. 

\begin{figure}
    \centering
    \includegraphics[width=1\linewidth]{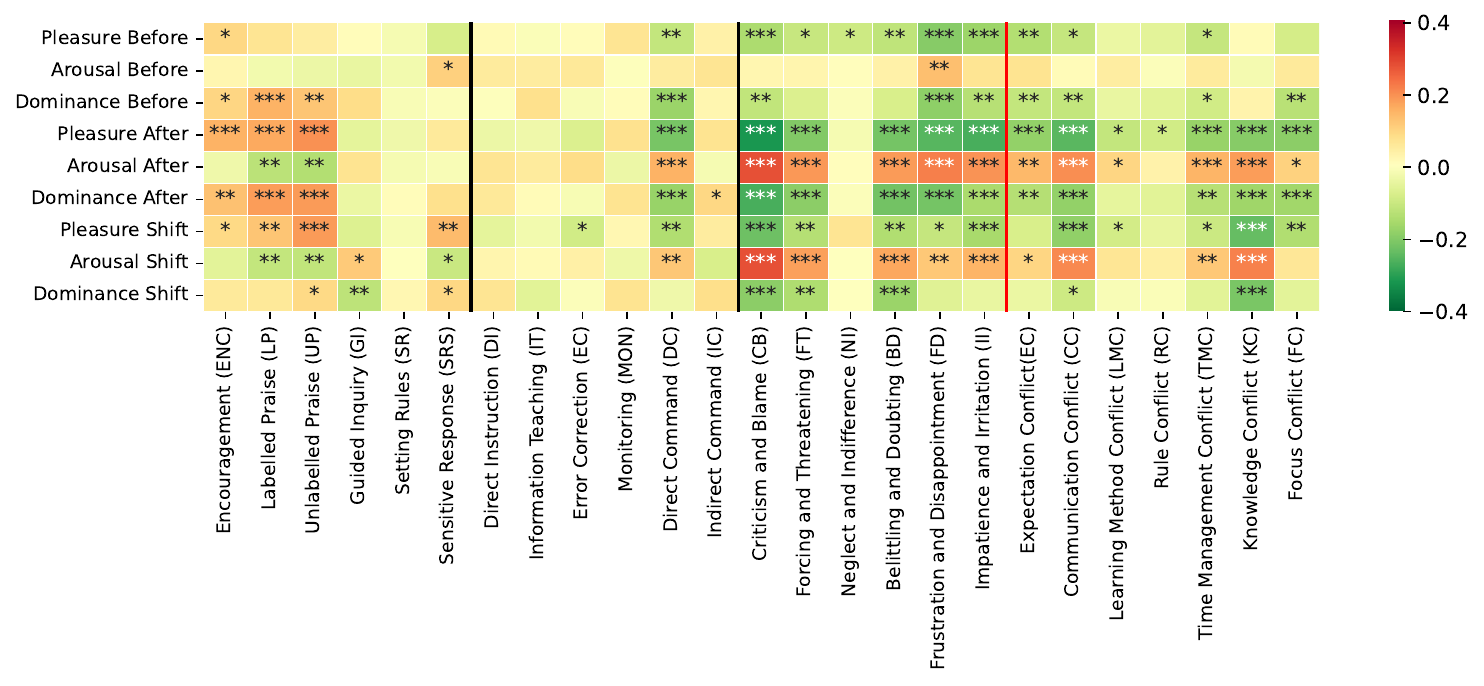}
    \caption{Correlation between emotions, behaviours and parent-child conflicts (* p<0.05, ** p<0.01, *** p<0.001).}
    \label{fig:heatmap:emotion_conflict_behaviours}
\end{figure}

Figure \ref{fig:heatmap:emotion_conflict_behaviours} illustrates the correlation between perceived emotions, parental behaviours, and parent-child conflicts. The bold lines divide behaviours into positive, neutral, and negative categories, while the red line separates behaviours from conflicts.  While correlation does not imply causation, it is valuable to examine how emotions before and after homework relate to behaviours and conflicts. 
As shown, greater pre-homework pleasure correlates positively with supportive behaviours like \textit{Encouragement}, and negatively with negative behaviours such as \textit{Criticism and Blame}, \textit{Forcing and Threatening}, and \textit{Belittling and Doubting}. Similarly, higher perceived dominance (a sense of control) before homework is associated with more positive behaviours and fewer conflicts, such as \textit{Expectation Conflict}, \textit{Communication Conflict}, and \textit{Time Management Conflict}.
This suggests that parents who start homework in a positive emotional state are more constructive and less likely to experience conflict.

%Post-homework emotions follow similar trends. Supportive behaviours are associated with higher post-session pleasure and dominance, and lower arousal—reflecting satisfaction and calm. In contrast, most negative behaviours correlate with decreased pleasure and dominance, and increased arousal, suggesting lingering emotional agitation. Notably, \textit{Neglect and Indifference} again shows no correlation, possibly reflecting disengagement.

%Post-homework emotions show a similar trend. Positive behaviours such as \textit{Encouragement} and \textit{Praise} are associated with higher pleasure and dominance, and lower arousal after homework sessions.  This pattern reflects everyday experiences: parents who engage in supportive behaviours often feel more satisfied and calm after homework sessions. In contrast, five out of six negative behaviours are significantly negatively correlated with post-homework pleasure and dominance, and positively correlated with arousal, indicating that negative behaviours tend to leave parents feeling more emotionally agitated and less in control after the session. \textit{Neglect and Indifference}, however, shows no significant correlation with post-homework emotions, suggesting that it may reflect emotional disengagement or detachment during the session.

Post-homework emotions follow similar trends. Positive behaviours are associated with higher post-homework pleasure and dominance, and lower arousal, reflecting satisfaction and calm. In contrast, most negative behaviours correlate with decreased post-homework pleasure and dominance, and increased arousal, indicating that negative behaviours tend to leave parents feeling more emotionally agitated and less in control after the session. Notably, \textit{Neglect and Indifference} again shows no correlation, possibly reflecting disengagement or detachment during the session.

Regarding conflicts, all conflict types negatively correlate with post-homework pleasure, aligning with Pekrun’s control-value theory of achievement emotions \cite{pekrun2006control}. Interestingly, \textit{Rule Conflict} shows the weakest association with post-homework emotions, indicating that it may be perceived as more neutral or procedural, potentially evoking fewer emotional reactions compared to other types of conflicts. As noted by Dumont et al. \cite{dumont2014quality}, disagreements over rules may represent normative aspects of homework interactions rather than deep interpersonal friction.
By contrast, \textit{Communication Conflict} and \textit{Knowledge Conflict} exhibit strong correlations with all post-homework emotions as well as with emotional shifts. These conflict types may be especially emotionally destabilising. Communication conflicts involve misunderstandings and frustration, while knowledge conflicts could highlight gaps in understanding between parents and children, triggering what Nunez et al. \cite{nunez2015relationships} describe as parental anxiety about academic competence. Both are associated with notable shifts in arousal (indicative of stress) and dominance (control), pointing to their potential to cause significant emotional upheaval.

In summary, these findings highlight the complex interplay among emotions, behaviours, and conflicts. While positive emotional states before homework are associated with more positive behaviours and fewer conflicts, emotionally charged dynamics (particularly involving communication or knowledge) can still arise. Moreover, conflict during homework appears to have a lasting emotional impact on parents, underscoring the importance of managing both how and when parental involvement occurs.

\section{Implications and Limitations}
\label{sec:limit}
In this study, we developed an LLM-based pipeline to analyse large-scale parent-child homework interactions, identifying parental behaviours and conflicts, and offering insights into parents' emotion dynamics during homework involvement. The findings carry several implications:

\textit{LLM-Assisted Analysis in Human-Centred Domains}.
While our study focuses on parent-child homework interactions, the proposed LLM-assisted coding framework is adaptable to other contexts involving complex interpersonal communication, such as healthcare consultations, workplace mentoring, and classroom discourse. To apply it elsewhere, one should: (1) define a new analytical target (e.g., support types, conflict, emotion); (2) collect relevant transcripts or recordings; and (3) iteratively construct a domain-specific codebook with expert input, guided by open, axial, and selective coding. GPT-4o can be effectively steered through prompt engineering and few-shot examples from expert-annotated data. Our modular pipeline also allows domain-specific adaptation of conflict dimensions (e.g., “trigger,” “intensity”), and future work may explore prompt or codebook transfer to accelerate deployment in related fields.

\textit{Implications for Family Education and Psychological Practice}.
By categorising parental behaviours and their emotional dynamics, this study enhances our understanding of how specific parenting behaviours influence child development and family relationships. The findings suggest that even seemingly supportive behaviours (e.g., unlabelled praise) might unintentionally contribute to conflict, allowing for more mindful interventions \cite{baumrind1991influence}. Furthermore, the emotional dynamics provide valuable insights into the psychological processes involved in parenting, which could lead to the development of more psychologically informed frameworks for family education \cite{denham1998emotional}. This research bridges gaps in understanding emotional and behavioural patterns, which is essential for crafting interventions that foster healthier, more supportive parent-child relationships.

\textit{Implications for Future Intervention Technologies}.
The emotional profiles captured in our study provide a foundation for designing emotion-sensitive tools to assist parents in managing homework-related stress. 
For example, systems could monitor emotional states (e.g., pleasure, dominance) and suggest adaptive strategies (e.g., reducing pressure, using calming techniques) during moments of tension. 
This aligns with the principles of \textit{Authoritative Parenting} \cite{gray1999unpacking}, which balances responsiveness and appropriate demands for positive child outcomes. Future technologies might provide tailored feedback to support parents in dynamically adjusting their strategies, such as offering autonomy when stress levels rise or recommending calming techniques during tense moments, ultimately promoting healthier parent-child relationships and supporting more effective parenting.

To the best of our knowledge, this is the first study to comprehensively investigate the emotional and behavioural dynamics of parental homework involvement through parent-child conversations. However, several limitations should be acknowledged:

\begin{figure}
    \centering
    \includegraphics[width=0.9
    \textwidth]{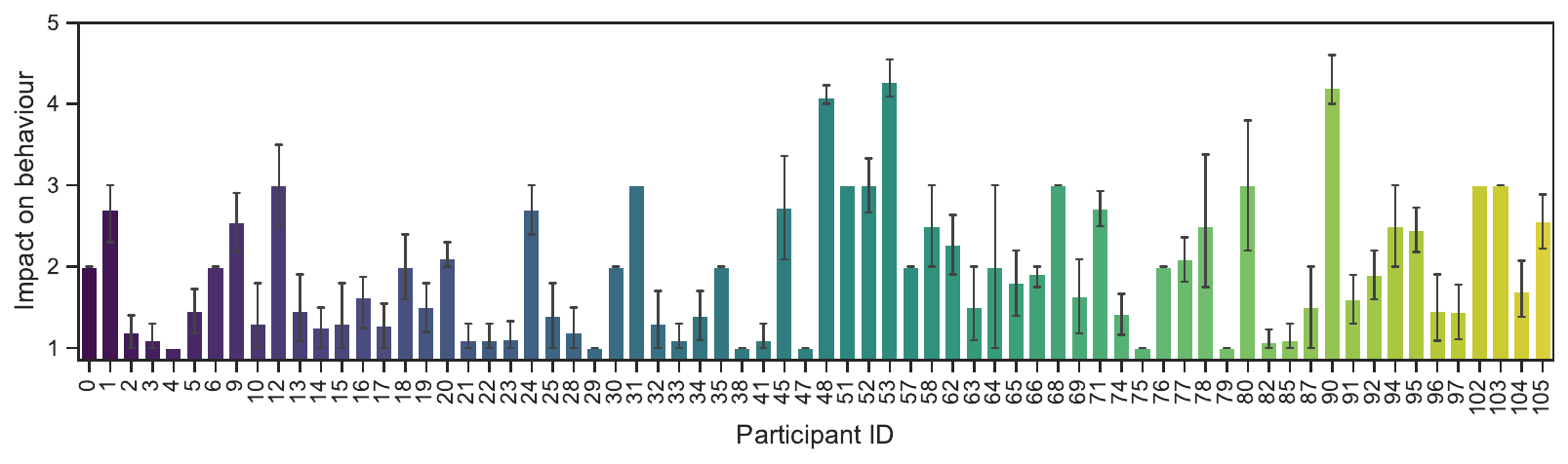}
    \caption{Different impacts of recording on educational behaviours (self-reported).}
    \label{fig:hawthorn}
\end{figure}

\textit{Sampling Bias}. As outlined in Section \ref{subsec: participants}, the parents in our sample had above-average education levels compared to national statistics in China. This creates a sampling bias and limits the generalisability of our findings. Future studies should aim for more diverse samples to ensure broader applicability.

\textit{Use of Text Transcripts Only}. Our analysis relied exclusively on text transcripts, omitting prosodic features such as tone, pitch, and speech dynamics. This decision protected participant privacy and simplified data processing, but may have reduced the granularity of emotional and behavioural analysis. Emotional nuances, such as sarcasm, irritation, or affection, are often conveyed through intonation rather than word choice. Text alone may also obscure the affective meaning of simple expressions like “oh” or “hmm,” which can vary widely depending on vocal delivery. Audio-based analysis could provide richer insight into conflict intensity and emotional expression even when the verbal content remains the same.

\textit{Hawthorne Effect}. The act of recording may have influenced participants' behaviour, a phenomenon known as the \textit{Hawthorne Effect}. To assess this, we asked parents to rate daily whether recordings affected their natural performance \footnote{The question is \textit{``Due to the fact that you knew the homework involvement behaviour today was recorded, did this affect your true performance?''}} using a 5-point Likert scale where 1 to 5 indicates \textit{``Completely unaffected''}, \textit{``Basically unaffected'}, \textit{``Slightly affected''}, \textit{``Significantly affected''} and \textit{`Extremely affected''}. As shown in Figure \ref{fig:hawthorn}, most participants (78.04\%) fell into \textit{'Completely unaffected'} or \textit{'Basically unaffected'} categories, with only 7.26\% reporting significant influence. While this suggests limited impact, we cannot fully rule out behavioural changes due to the presence of recording.

%The act of recording may have influenced participants' behaviour—a phenomenon known as the Hawthorne Effect. To assess this, we asked parents to rate daily whether recording affected their natural performance using a 5-point Likert scale\footnote{“Due to the fact that you knew the homework involvement behaviour today was recorded, did this affect your true performance?”}. As shown in Figure \ref{fig:hawthorn}, 78.04% of responses fell into “Completely unaffected” or “Basically unaffected” categories, while only 7.26% reported significant influence. While this suggests limited impact, we cannot fully rule out behavioural changes due to the presence of recording.

\textit{Privacy Concerns}. Privacy is a critical consideration in our research. We took extensive measures to ensure that participants were fully informed about the data collection and future usage, and strictly limited the scope of our study to homework-related interactions. Data was stored on secure hardware to protect the participants’ privacy. %However, using transcripts instead of more detailed audio-visual data due to privacy concerns may have constrained the depth of our analysis.
However, to protect privacy, we opted not to collect audio-visual data, which may have constrained the depth and accuracy of affective and behavioural analysis.

\section{Conclusion}

While parental homework involvement is often seen as beneficial, it can also be a significant source of stress and conflict.
Despite this complexity, educational systems frequently assume that parents can provide effective support without sufficient guidance, overlooking the emotional and relational challenges that homework interactions entail. This study addresses a key methodological gap by demonstrating how large-scale, naturalistic parent-child interactions can be unobtrusively analysed through LLM-powered conversation analysis. 
Drawing on 475 hours of in-home audio recordings from 78 Chinese families, we reveal nuanced emotional and behavioural dynamics that traditional retrospective or observational methods may overlook. Our analysis uncovers how parents’ emotional states shift throughout the homework process and how these shifts intersect with specific behavioural patterns and conflict types.

Importantly, our findings challenge the assumption that “positive” behaviours always yield positive outcomes. Even well-intentioned actions such as \textit{Unlabelled Praise} can be linked to increased conflict. Beyond its methodological contributions, this work enriches theoretical understanding of family education by illuminating the emotional texture of everyday parent-child exchanges. These insights have implications not only for research in ubiquitous computing, computational social science, and education, but also for designing more empathetic, evidence-based interventions that support families in navigating the complexities of homework involvement.

%%
%% The acknowledgments section is defined using the "acks" environment
%% (and NOT an unnumbered section). This ensures the proper
%% identification of the section in the article metadata, and the
%% consistent spelling of the heading.
\begin{acks}
This work is supported by the Natural Science Foundation of China (Grant No. 62302252).

\end{acks}

%%
%% The next two lines define the bibliography style to be used, and
%% the bibliography file.
\bibliographystyle{ACM-Reference-Format}
\bibliography{refs}

\appendix
%TC:ignore
\section*{Appendix}
\section{More Information about Collected Data}
\begin{figure}
    \centering
    \includegraphics[width=0.99\linewidth]{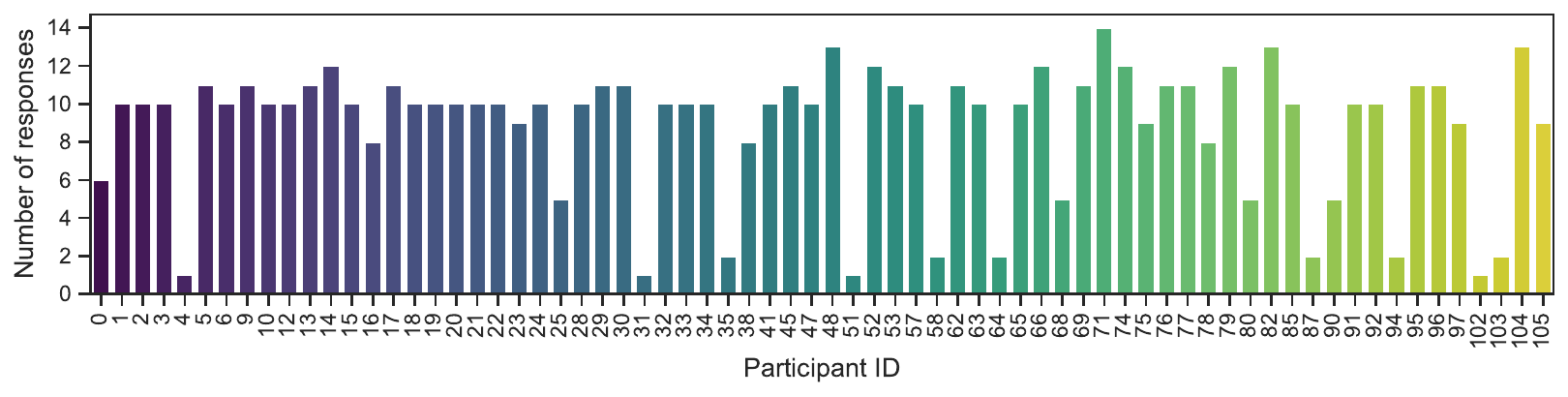}
    \caption{\added{Number of responses for different participants}.}
    \label{fig:valence_count}
\end{figure}

Figure \ref{fig:valence_count} shows the number of responses (daily surveys paired with audio recordings) for each participant. 

\begin{table}
\centering
\footnotesize
\caption{Coding usage guidelines for parent behaviour during homework involvement.}
\label{appen:tab:behaviour}
\begin{tabular}{p{0.22\textwidth} p{0.7\textwidth}}
\toprule
\textbf{Code Name} & \textbf{Usage Guidelines} \\ \midrule
\textit{Encouragement (ENC)} & Use this code when parents offer positive reinforcement and encouragement, especially when the child faces difficulties, regardless of the outcome. \\ \hline
\textit{Labelled Praise (LP)} & Use this code when parents clearly point out a specific behaviour, achievement, or performance and give positive feedback. \\ \hline
\textit{Unlabelled Praise (UP)} & Use this code when parents give general, unspecific praise without focusing on particular details or behaviours. \\ \hline
\textit{Guided Inquiry (GI)} & Use this code when parents help the child find a solution by asking questions or offering hints, rather than directly giving the answer. \\ \hline
\textit{Setting Rules (SR)} & Use this code when parents establish clear rules and expect the child to follow them. These rules are often related to homework order, time, or completion standards. \\ \hline
\textit{Sensitive Response (SRS)} & Use this code when parents recognize the child's emotions and provide comfort or emotional support. \\ \hline
\textit{Direct Instruction (DI)} & Use this code when parents provide direct answers or solutions without encouraging the child to think through the problem. \\ \hline
\textit{Information Teaching (IT)} & Use this code when parents provide systematic explanations to help the child understand new knowledge or concepts. \\ \hline
\textit{Error Correction (EC)} & Use this code when parents identify errors in the child's homework and guide them to make corrections. \\ \hline
\textit{Monitoring (MON)} & Use this code when parents monitor the child's homework progress or check the quality of their work. \\ \hline
\textit{Direct Command (DC)} & Use this code when parents give a firm command in a direct and authoritative manner. \\ \hline
\textit{Indirect Command (IC)} & Use this code when parents use a more subtle or indirect approach to encourage the child to complete tasks. \\ \hline
\textit{Criticism \& Blame (CB)} & Use this code when parents criticize or blame the child for not meeting expectations. \\ \hline
\textit{Forcing \& Threatening (FT)} & Use this code when parents use threats or force to make the child complete a task. \\ \hline
\textit{Neglect \& Indifference (NI)} & Use this code when parents ignore or display indifference toward the child's needs, emotions, or requests for attention. \\ \hline
\textit{Belittling \& Doubting (BD)} & Use this code when parents use belittling or doubting language to undermine the child's abilities directly. \\ \hline
\textit{Frustration \& Disappointment (FD)} & Use this code when parents express disappointment due to the child's academic performance or lack of progress. \\ \hline
\textit{Impatience \& Irritation (II)} & Use this code when parents become frustrated or impatient with the child's learning pace or performance. \\ \bottomrule
\end{tabular}
\end{table}

\begin{table}
\centering
\footnotesize
\caption{Coding usage guidelines for parent-child conflict during homework involvement.}
\label{appen:tab:conflict}
\begin{tabular}
{p{0.12\textwidth} p{0.83\textwidth}}

\toprule
\textbf{Code Name} & \textbf{Usage Guidelines} \\ 
\midrule

\textit{Expectation \newline Conflict} & Apply this code when there is a significant discrepancy between parents' expectations and the child’s self-perception or goals. This may also include cases where parents compare the child to peers or siblings, adding pressure. \\

\hline
\textit{Communication Conflict} & Use this code when communication breakdowns occur due to negative communication styles such as parental criticism or questioning, leading to frustration. Pay attention to whether the parent blames or belittles the child and the child’s emotional response to it. \\

\hline
\textit{Learning Method Conflict} & Use this code when parents attempt to enforce a change in the child’s learning method or directly intervene, particularly when this disagreement leads to conflict. \\

\hline
\textit{Rule Conflict} & Use this code when the conflict revolves around parental rules, boundaries, or control, such as how and when the child should complete their homework. Pay special attention to cases where the child expresses dissatisfaction with the rules or seeks more autonomy. \\

\hline
\textit{Time \newline Management Conflict} & Use this code when the conflict centers on how study time is managed, the frequency of study sessions, or the child’s energy allocation. This includes situations where the child expresses dissatisfaction with the time constraints imposed by parents. \\

\hline
\textit{Knowledge \newline Conflict} & Apply this code when the conflict stems from a difference in knowledge mastery, a lack of understanding of the child’s learning difficulties, or when the child questions the parent’s knowledge. Pay special attention to instances where the parent underestimates the difficulty of the material for the child. \\

\hline
\textit{Focus Conflict} & Use this code when the conflict is driven by the parent’s dissatisfaction with the child’s focus or attention during study, particularly when repeated parental intervention causes tension. \\

\bottomrule
\end{tabular}
\end{table}

\section{Coding Usage Guidelines for Homework Assistance\label{sec:guide}}

This section outlines the coding guidelines used for analysing parent-child conflicts and parental behaviours during homework assistance (see Tables \ref{appen:tab:behaviour} and \ref{appen:tab:conflict}). The coding manual ensures consistency in annotation by providing clear definitions, usage examples, and contextual notes for each code. These guidelines serve as a practical reference for researchers and coders, supporting systematic identification and categorisation of key behaviours and conflict types. Each code is accompanied by detailed instructions on its application to promote reliability and analytical rigor.

\section{Analysis of Category Confusion in GPT-4o Coding Using Confusion Matrix\label{sec:Confusion}}

To assess the alignment between GPT-4o and expert annotations, we used confusion matrices to examine coding performance across behavioural and conflict categories. These matrices visualise which categories GPT-4o correctly identified and which it misclassified, offering insights into systematic biases in GPT-4o’s performance, with certain categories being more prone to misclassification.

\begin{figure}
    \centering
    \includegraphics[width=0.65\linewidth]{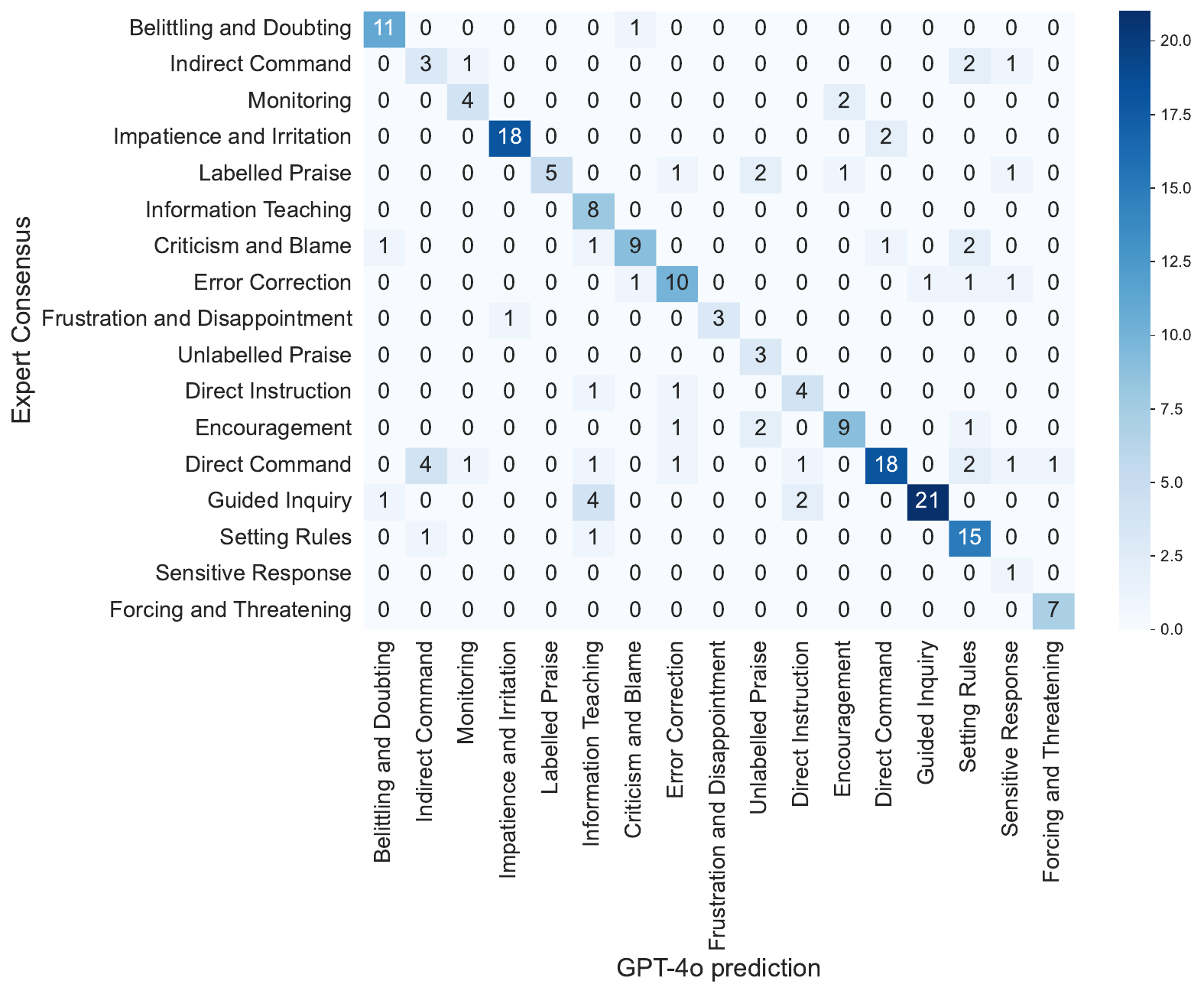}
    \caption{Confusion matrix of parental behaviour classification by GPT-4o. Diagonal cells indicate correct classifications, and off-diagonal cells reflect misclassifications.}
    \label{fig:behaviour_confusion}
\end{figure}

\begin{figure}
    \centering
    \includegraphics[width=0.6\linewidth]{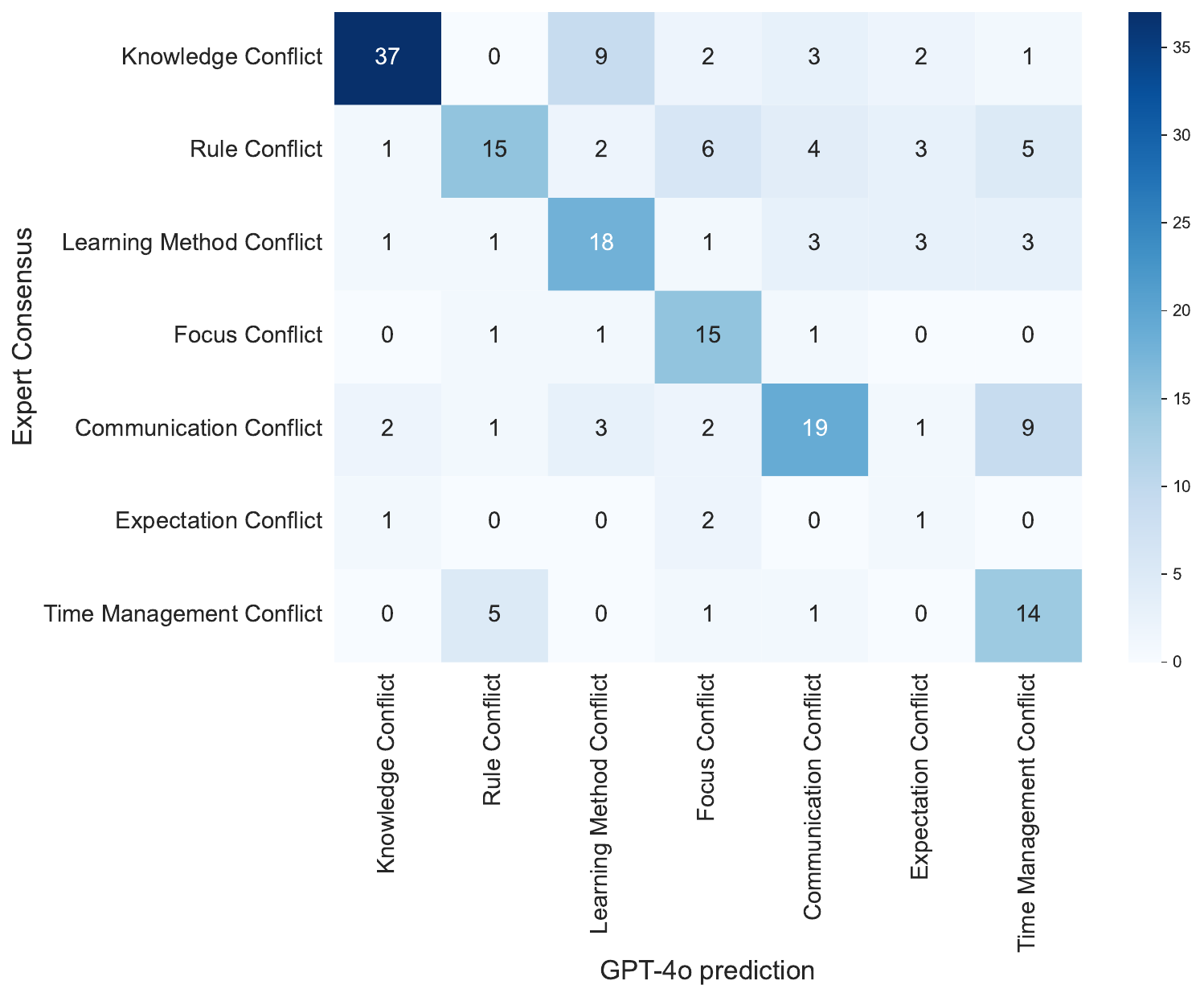}
    \caption{Confusion matrix of parent-child conflict classification by GPT-4o. Diagonal cells indicate correct matches and off-diagonal cells reflect areas of confusion.}
    \label{fig:conflict_confusion}
\end{figure}

%From the parental behaviour confusion matrix, we can observe that GPT-4o performed well in certain categories, such as \textit{Information Teaching} (IT) (28 matches), \textit{Error Correction} (EC) (14 matches), and \textit{Direct Command} (DC) (19 matches), demonstrating strong consistency with the expert consensus. However, significant confusion was observed in categories like \textit{Indirect Command} (IC), which was frequently misclassified as \textit{Direct Command} (DC) and \textit{Guided Inquiry} (GI), suggesting difficulties in distinguishing between direct and indirect commands. Additionally, \textit{Sensitive Response} (SRS) and \textit{Impatience and Irritation} (II) were often confused with \textit{Encouragement} (ENC) and \textit{Monitoring} (MON), indicating challenges in recognizing emotional support behaviours. GPT-4o also struggled with emotional categories such as \textit{Neglect and Indifference} (NI) and \textit{Frustration and Disappointment} (FD), as well as forceful behaviours like \textit{Forcing and Threatening} (FT) and \textit{Criticism and Blame} (CB), where misclassifications were frequent, as shown in Figure \ref{fig:behaviour_confusion}.

\textit{Behaviour Confusion Analysis}. As shown in Figure \ref{fig:behaviour_confusion}, GPT-4o demonstrated high consistency with expert annotations in structured behavioural categories, including \textit{Information Teaching} (28 matches), \textit{Error Correction} (14 matches), and \textit{Direct Command} (19 matches). However, several categories were frequently confused. For example, \textit{Indirect Command} (IC) was often misclassified as \textit{Direct Command} or \textit{Guided Inquiry}, suggesting difficulties in differentiating between directive styles. Emotional behaviours also posed challenges: \textit{Sensitive Response} and \textit{Impatience and Irritation} were frequently mistaken for \textit{Encouragement} and \textit{Monitoring}, indicating ambiguity in identifying emotional tone. Misclassifications were also observed in affectively charged categories such as \textit{Neglect and Indifference}, \textit{Frustration and Disappointment}, \textit{Forcing and Threatening}, and \textit{Criticism and Blame}.

\textit{Conflict Confusion Analysis}. %Similarly, from the parent-child conflict confusion matrix, GPT-4o showed strong performance in structured categories such as \textit{Knowledge Conflict} (KC) (26 matches), \textit{Time Management Conflict} (TMC) (19 matches), and \textit{Focus Conflict} (FC) (16 matches). However, frequent confusion occurred between \textit{Expectation Conflict} (EC) and \textit{Knowledge Conflict} (KC) (11 misclassifications), as well as between \textit{Learning Method Conflict} (LMC) and \textit{Time Management Conflict} (TMC), suggesting overlap in parental interventions. Emotional and communication-based categories, such as \textit{Emotional Conflict} (EMC) and \textit{Communication Conflict} (CC), were also often misclassified, reflecting difficulties in distinguishing between these nuanced conflicts. \textit{Rule Conflict} (RC) was another challenging category, with misclassifications occurring in relation to \textit{Time Management Conflict} (TMC) and \textit{Learning Method Conflict} (LMC), as illustrated in Figure \ref{fig:conflict_confusion}.
Similarly, Figure \ref{fig:conflict_confusion} shows that GPT-4o performed well in coding structured conflict types like \textit{Knowledge Conflict} (26 matches), \textit{Time Management Conflict} (19 matches), and \textit{Focus Conflict} (16 matches). However, overlaps were observed in conceptually adjacent categories. For instance, \textit{Expectation Conflict} was often confused with \textit{Knowledge Conflict}, and \textit{Learning Method Conflict} was misclassified as \textit{Time Management Conflict}, suggesting blurred boundaries in parental interventions. Greater confusion occurred in emotionally and interpersonally complex categories. \textit{Emotional Conflict} and \textit{Communication Conflict} were frequently misclassified, highlighting GPT-4o’s limitations in parsing subtle relational tensions. \textit{Rule Conflict} was another challenging category, with misclassification occurring in relation to \textit{Time Management Conflict} and \textit{Learning Method Conflict}, possibly due to overlaps in rule enforcement and learning structure.

In conclusion, the confusion matrices indicate that GPT-4o performs reliably in well-defined, instructional categories but faces difficulties with emotionally nuanced and context-dependent behaviours and conflicts. Enhancing GPT-4o's sensitivity to emotional tone, implicit meaning, and contextual cues may significantly improve its performance in these areas, offering more accurate and human-aligned coding in future analyses.

\section{Prompt for Data Processing}

\subsection{Prompt for Fixing Transcription Errors}
\label{app:prompt_transcription}

\begingroup
\footnotesize
\begin{spverbatim}
# Correcting Transcription Errors in Parental Homework Involvement (Python Dictionary Format)

## Objective
Correct transcription errors in the text of parents helping their kids with homework while preserving the original expression style.

## Input
- A list of Python dictionaries, each containing:
  - Conversation ID (`id`)
  - Speaker (`speaker`)
  - Transcribed content (`content`)

## Processing Steps
1. **Parse the Dictionary List**: Read through each transcription entry.
2. **Identify Transcription Errors**: Carefully examine and identify homophones or typographical errors in the `content` section.
3. **Correct Errors**: Correct the errors without altering the original sentence style to improve the clarity and accuracy of the text.
4. **Maintain Original Style**: When updating the `content` field, ensure that the original expression style is preserved, striving to maintain the natural flow of the original text.
5. **Structure Validation**: After correction, ensure that the structure of each dictionary remains identical to the corresponding original dictionary, including but not limited to field order and data types.

## Output Format
Please return the output in the following strict format, retaining only the `id` and `content` fields and correcting only the `content` field:
```python
[
    {'id': 1, 'content': 'Corrected content 1'}, 
    {'id': 2, 'content': 'Corrected content 2'}, 
    {'id': 3, 'content': 'Corrected content 3'}, 
    ...
]

## Considerations
- **Retain Conversational Style**: Pay special attention to preserving the conversational characteristics of the transcribed text when correcting errors.
- **Minimize Intervention**: Adjustments should minimally impact the original meaning and expression, correcting only obvious typographical errors and homophone transcription errors while avoiding over-editing.
- **Clarity and Accuracy**: While retaining the conversational style, ensure that the clarity and accuracy of the text are enhanced.
- **Structural Consistency**: Ensure the output dictionary structure is fully consistent with the input data, with the same number of entries, field order, and data types, and only the `content` field is corrected as needed.
- **Length Consistency**: Ensure that each dictionary's length after correction matches the original data, avoiding any length changes due to content addition or omission.
- **Preserve All Entries**: Ensure all original entries are retained in the output data, with no omissions or deletions.
- **No Deletion of Original Content**: Ensure all original content is preserved, with corrections made without deleting any parts, especially numbers and sentences.

\end{spverbatim}
\endgroup

\subsection{Prompt for Role Recognition from Transcripts}
\label{app: prompt_role recognition}
\begingroup
\footnotesize
\begin{spverbatim}
# Speaker Role Identification Task

## Task Description
Use transcribed audio text to identify the roles of speakers in the context of parents helping their children with homework. This API can automatically identify the number of speakers and assign each segment of speech to a specific speaker (e.g., 'speaker 1', 'speaker 2', etc.). When two speakers are identified, it is usually assumed that one is the parent and the other is the child. If three or more speakers are identified, there might be inaccuracies in the number of speakers detected by the API. In such cases, two or more identified speakers might actually be the same person (e.g., both the parent or both the child) or might include others (e.g., someone playing a recording of a lesson).

## Input Format
The input is a list of dictionaries, each corresponding to a segment of transcribed text. Each dictionary contains the following keys:
- "id": The conversation ID.
- "speaker": A string representing the label assigned to the speaker (e.g., "speaker 1").
- "content": A string containing the transcribed text for that segment.

## Output
The output should be a dictionary mapping speaker labels to their inferred roles. The roles can be "parent", "child", or "others". The "others" role should only be used when there are three or more speakers.

### Role Inference Guidelines
- **parent**: Typically asks questions, assigns tasks, gives instructions, commands, explains concepts, criticizes, encourages, or provides help.
- **child**: Typically requests help/guidance/praise, expresses confusion, or asks questions.
- **others**: Recordings that appear in the conversation can be labeled as "others". "others" do not participate in the conversation and typically do not respond to "parent" or "child".

### Output Format
Please return the result in the following format, strictly as a dictionary without explanatory text:
```python
{
    "Speaker 1": "",
    "Speaker 2": "",
    ...
}

### Considerations
1. Role inference should be based on the content of the conversation, not just the sequence of speakers.
2. For three or more speakers, carefully distinguish between roles. Only label as "others" when it is very clear; otherwise, do not infer "others".
3. For each speaker, analyze their speech content in detail to avoid mistakenly labeling an actual participant in the conversation as "others". Consider marking "others" only in very rare cases.

\end{spverbatim}
\endgroup
\normalsize

\section{Prompt for Understanding Behavioural Dynamics During Homework Involvement}

\normalsize
\subsection{Prompt for Parent Behaviour Identification}

\begingroup
\footnotesize
\begin{spverbatim}
# Task Description
You are a family education expert, and your task is to analyze the specific behaviors of parents during homework assistance. You need to carefully review the provided dialogue content, identify and capture dialogue segments that reflect parental behavior, and classify these behaviors using predefined behavior categories.

## Behavior Categories

### Positive Behaviors

#### 1. **Encouragement (ENC)**
- **Code Definition**: The parent actively supports the child's efforts and progress through words or actions, boosting the child's confidence and motivation to learn, helping them overcome difficulties.
- **Usage Guidelines**: Use when the parent provides positive support and encouragement, especially when the child encounters difficulties, and the parent maintains a positive attitude regardless of the outcome.
- **Example**:  
  - Parent: "You've worked really hard, keep it up! I believe in you!"
  - Parent: "Don’t worry, we’ll take it step by step, you’ll definitely get it."

#### 2. **Labelled Praise (LP)**
- **Code Definition**: The parent explicitly points out specific behaviors or achievements of the child and offers praise, helping the child recognize their concrete progress and strengths.
- **Usage Guidelines**: Use when the parent clearly identifies a specific behavior, achievement, or performance of the child and provides positive feedback.
- **Example**:  
  - Parent: "You did this addition problem perfectly, no mistakes at all!"
  - Parent: "Your handwriting is really neat this time, keep it up!"

#### 3. **Unlabelled Praise (UP)**
- **Code Definition**: The parent offers general praise to the child but does not point out any specific behaviour or achievement.
- **Usage Guidelines**: Use when the parent gives unlabelled praise without mentioning specific details or actions.
- **Example**:  
  - Parent: "You're doing great, keep going!"
  - Parent: "Wow, amazing!"

#### 4. **Guided Inquiry (GI)**
- **Code Definition**: The parent guides the child to think independently and solve problems through questions or hints, rather than directly providing the answer, encouraging the child to explore and think critically.
- **Usage Guidelines**: Use when the parent helps the child find solutions through questions or hints rather than directly giving the answer.
- **Example**:  
  - Parent: "Where do you think this letter should go?"
  - Parent: "What strategy can we use to solve this problem? Think about a few ways."

#### 5. **Setting Rules (SR)**
- **Code Definition**: The parent sets clear rules or requirements for the child to complete their homework, helping the child establish good study habits and time management skills.
- **Usage Guidelines**: Use when the parent sets specific rules and requires the child to follow them, typically related to the order of tasks, time, or completion standards.
- **Example**:  
  - Parent: "You need to finish your language homework first before you can watch cartoons."
  - Parent: "You have to finish all your homework before dinner, then you can go out to play."

#### 6. **Sensitive Response (SRS)**
- **Code Definition**: The parent responds to the child's emotions, needs, and behaviors in a timely, appropriate, and empathetic manner. The parent perceives the child's feelings and provides emotional support.
- **Usage Guidelines**: Use when the parent recognizes the child’s emotions and offers understanding and emotional comfort or support.
- **Example**:  
  - Parent: "I know you're feeling tired, let's take a break and continue later, okay?"
  - Parent: "Are you finding this problem difficult? It’s okay, we’ll go over it together."

### Neutral Behaviors

#### 7. **Direct Instruction (DI)**
- **Code Definition**: The parent directly tells the child how to complete a task or solve a problem without using an exploratory or guided approach.
- **Usage Guidelines**: Use when the parent provides the answer or solution directly without guiding the child to think through questions or hints.
- **Example**:  
  - Parent: "You should do it like this, add 4 to 6, and it equals 10."
  - Parent: "Just copy the answer directly, don’t overthink it."

#### 8. **Information Teaching (IT)**
- **Code Definition**: The parent imparts new knowledge or skills to the child through explanations, such as explaining concepts or lessons, to help the child understand new study material.
- **Usage Guidelines**: Use when the parent provides systematic explanations to help the child learn new knowledge or skills.
- **Example**:  
  - Parent: "The word ‘tree’ is written with a wood radical on the left and ‘inch’ on the right, let’s write it together."
  - Parent: "The multiplication table goes like this, two times two equals four, two times three equals six. Let’s memorize these first."

#### 9. **Error Correction (EC)**
- **Code Definition**: The parent points out mistakes in the child’s homework and guides them to make corrections.
- **Usage Guidelines**: Use when the parent identifies mistakes in the child’s homework and guides the child to correct them.
- **Example**:  
  - Parent: "You missed the ‘wood’ radical here, write it again."
  - Parent: "This addition is wrong, redo it, and make sure to align the columns."

#### 10. **Monitoring (MON)**
- **Code Definition**: The parent periodically checks the child’s homework progress or completion to ensure tasks are done on time.
- **Usage Guidelines**: Use when the parent checks the child’s homework progress or reviews the quality of the work.
- **Example**:  
  - Parent: "How many pages have you written? Let me check for mistakes."
  - Parent: "I’ll check your pinyin homework today to see if there are any issues."

#### 11. **Direct Command (DC)**
- **Code Definition**: The parent uses clear, direct language to command the child to perform a task or behavior, often with a strong, imperative tone.
- **Usage Guidelines**: Use when the parent gives a command in a forceful manner, requiring the child to complete a task.
- **Example**:  
  - Parent: "Go do your math homework right now, no more delays!"
  - Parent: "Stop playing with your toys and finish copying your pinyin."

#### 12. **Indirect Command (IC)**
- **Code Definition**: The parent uses a more indirect approach to convey a request to the child, such as a suggestion or implication, rather than giving a direct order.
- **Usage Guidelines**: Use when the parent suggests or implies that the child should complete a task without directly commanding them.
- **Example**:  
  - Parent: "Have you finished your homework? Maybe it's time to get it done?"
  - Parent: "Let’s finish the homework first, so we don’t have to worry about it later."

### Negative Behaviors

#### 13. **Criticism & Blame (CB)**
- **Code Definition**: The parent expresses negative evaluations of the child’s mistakes or behavior, directly blaming the child for their shortcomings, potentially using belittling language.
- **Usage Guidelines**: Use when the parent criticizes or blames the child for not meeting expectations.
- **Example**:  
  - Parent: "How could you get such an easy character wrong?"
  - Parent: "I’ve told you a thousand times, why can't you remember?"

#### 14. **Forcing & Threatening (FT)**
- **Code Definition**: The parent applies pressure or threatens consequences to force the child to comply with their demands, aiming to achieve the desired behavior.
- **Usage Guidelines**: Use when the parent uses threats or force to demand task completion.
- **Example**:  
  - Parent: "If you don’t finish your homework, you won’t get to play with your blocks!"
  - Parent: "If you don’t finish it, I’ll take away your toys!"

#### 15. **Neglect & Indifference (NI)**
- **Code Definition**: The parent shows disregard or indifference to the child’s needs or emotions, providing no attention or response.
- **Usage Guidelines**: Use when the parent ignores the child’s requests, emotions, or behavior, showing no care or reaction.
- **Example**:  
  - Child: "Mom, can you help me with this problem?"
  - Parent (ignores the child and continues using the phone).

#### 16. **Belittling & Doubting (BD)**
- **Code Definition**: The parent belittles or doubts the child’s abilities, directly undermining the child’s confidence and motivation. This type of language or behavior often conveys distrust or dissatisfaction with the child’s abilities.
- **Usage Guidelines**: Use when the parent uses derogatory language or expresses doubt about the child’s abilities, typically showing a lack of confidence in the child’s learning capabilities or expressing strong dissatisfaction with their performance.
- **Example**:  
  - Parent: "How can you be so dumb, you can’t even do simple addition?"
  - Parent: "With grades like yours, there’s no way you’ll get into a good school."

#### 17. **Frustration & Disappointment (FD)**
- **Code Definition**: The parent expresses feelings of frustration or disappointment due to the child’s performance not meeting expectations, often accompanied by negative emotional expressions.
- **Usage Guidelines**: Use when the parent expresses disappointment due to the child’s academic performance or progress not meeting expectations.
- **Example**:  
  - Parent: "I didn’t expect you to do so poorly, you really let me down."
  - Parent: "I thought you would do better, I guess I was wrong."

#### 18. **Impatience & Irritation (II)**
- **Code Definition**: The parent shows impatience or irritation due to the child’s performance not meeting expectations, often accompanied by negative language or behavior.
- **Usage Guidelines**: Use when the parent expresses impatience or frustration with the child’s learning pace or performance.
- **Example**:  
  - Parent: "Why are you so slow? I’ve been waiting forever!"
  - Parent: "How come you still haven’t finished? You’re always dragging your feet!"

# Input Format
The input is a list of dictionaries representing audio segments, each dictionary contains:
- 'id': Identifier for the segment.
- 'speaker': The role of the speaker (parent, child, or others).
- 'content': The dialogue content of the segment.

# Output Format
The output is a JSON object, segmenting the dialogue record into different behaviors labeled with a "behaviour_id" (each behavior may consist of one or more sentences). The corresponding start and end IDs of the dialogue segment are marked for each behavior. Provide a description of the behavior in the "Description of behavior" and then map this behavior to the parent behavior code abbreviation "code," ensuring accurate classification.

The output should be strictly returned in the following format, enclosed in ```json```:

```json
[
    {
        "behaviour_id": 1,
        "Start ID": The start ID of the dialogue segment corresponding to this behavior,
        "End ID": The end ID of the dialogue segment corresponding to this behavior,
        "Description of behavior": "Description of the behavior",
        "Parent Behavior": "Specific words of the parent",
        "code": "Behavior abbreviation",
    }
    ...
]
```

# Example Output
```json
[
    {
        "behaviour_id": 1,
        "Start ID": The start ID of the dialogue segment corresponding to this behavior,
        "End ID": The end ID of the dialogue segment corresponding to this behavior,
        "Description of behavior": "The parent gave clear positive feedback on the child’s math performance.",
        "Parent Behavior": "You did a great job on your math today!",
        "code": "Labelled Praise (LP)"
    },
    {
        "behaviour_id": 2,
        "Start ID": The start ID of the dialogue segment corresponding to this behavior,
        "End ID": The end ID of the dialogue segment corresponding to this behavior,
        "Description of behavior": "The parent encouraged the child to keep trying and set a goal to reduce errors, expressing hope for future improvement.",
        "Parent Behavior": "Keep it up, let’s aim to make fewer mistakes next time!",
        "code": "Encouragement (ENC)"
    },
    {
        "behaviour_id": 3,
        "Start ID": The start ID of the dialogue segment corresponding to this behavior,
        "End ID": The end ID of the dialogue segment corresponding to this behavior,
        "Description of behavior": "The parent gave the answer directly without encouraging independent thinking.",
        "Parent Behavior": "The answer to this problem is 3, just write it down.",
        "code": "Direct Instruction (DI)"
    },
......
]
```
\end{spverbatim}
\endgroup
\normalsize

\subsection{Prompt for Parent-Child Conflict Identification}

\begingroup
\footnotesize
\begin{spverbatim}
# Parent-Child Conflict Identification Task

## Below are common classifications of parent-child conflicts during homework assistance:

### 1. **Expectation Conflict (EC)**
- **Code Definition**: Parents have high expectations for their child's academic performance, progress, or future, while the child's actual abilities, goals, or interests do not align with these expectations, leading to conflict. Parents may compare the child's performance with others, intensifying the conflict.
- **Usage Guidelines**: Use when parents have excessive demands on the child's performance or when there is a clear discrepancy between the parent's expectations and the child's self-perception. This includes situations where the parent compares the child to others (e.g., classmates, siblings), causing pressure.
- **Example**:  
  - Parent: "You should score full marks like your classmates. How can you make mistakes on such easy questions?"
  - Child: "I've done my best! Why do you always think I'm worse than others?"

### 2. **Communication Conflict (CC)**
- **Code Definition**: Conflict arising from different communication styles between parents and children during homework assistance. Parents may criticize, question, or belittle the child, leading the child to feel misunderstood or oppressed, worsening communication barriers.
- **Usage Guidelines**: Use when negative communication, such as emotional outbursts, criticism, or questioning, leads to conflict. Pay attention to whether the parent blames or belittles the child's performance and the child's resulting resistance.
- **Example**:  
  - Parent: "What's wrong with you? I've explained this so many times, and you still don't get it!"
  - Child: "I don't want to listen to you anymore! You always yell at me like this!"

### 3. **Learning Method Conflict (LMC)**
- **Code Definition**: Disagreements between parents and children on how to complete homework. Parents may feel the child's methods are inefficient and try to enforce their own, while the child insists on using their own approach and resists intervention.
- **Usage Guidelines**: Use when the parent tries to force the child to change learning methods or directly intervene in the learning process, especially when disagreements over methods lead to conflict.
- **Example**:  
  - Parent: "You can't study like this; you should finish all the questions first and then check them!"
  - Child: "This is how I like to do it! Why should I follow your way?"

### 4. **Rule Conflict (RC)**
- **Code Definition**: Conflict between parental control and the child's autonomy regarding rules set for studying. Parents may try to control the child's study schedule and methods through strict rules, while the child seeks more autonomy and flexibility.
- **Usage Guidelines**: Use when conflict involves rules, boundaries, or control set by parents, such as study time or homework completion methods. Pay attention to whether the child expresses dissatisfaction with the rules or seeks more autonomy.
- **Example**:  
  - Parent: "You must do your homework right after dinner; no more procrastinating!"
  - Child: "I want to play a little longer! You always control everything!"

### 5. **Time Management Conflict (TMC)**
- **Code Definition**: Disagreement between parents and children on how to allocate time and energy for studying. Parents may expect the child to follow a fixed schedule, while the child has a different time arrangement, leading to conflict.
- **Usage Guidelines**: Use when conflict revolves around study time, energy allocation, or the pace of study. Especially focus on whether the child expresses dissatisfaction with the parent's time requirements.
- **Example**:  
  - Parent: "You always leave your homework until late at night. Your efficiency is too low!"
  - Child: "I prefer studying later. I can't focus in the morning!"

### 6. **Knowledge Conflict (KC)**
- **Code Definition**: Conflict caused by differences in knowledge level or understanding between parents and children. Parents, having mastered certain knowledge, fail to understand why the child struggles or cannot explain things from the child’s perspective, leading the child to feel misunderstood or criticized. Additionally, parents' unfamiliarity with certain study content may cause the child to doubt the correctness of their guidance, further sparking conflict.
- **Usage Guidelines**: Use when conflict involves differences in knowledge mastery, the parent's lack of understanding of the child's learning difficulties, or the child's doubt in the parent's knowledge. Pay attention to whether parents underestimate the difficulty of the child's learning.
- **Example**:  
  - Parent: "This problem is so simple; how can you not get it?"
  - Child: "What you’re explaining is different from what the teacher said. I don’t understand."

### 7. **Focus Conflict (FC)**
- **Code Definition**: Parents are dissatisfied with the child's focus during the study, believing the child is distracted or inefficient. They try to remind or correct the child to stay focused, while the child may feel pressured by excessive interference, resulting in conflict.
- **Usage Guidelines**: Use when conflict arises because the parent feels the child is not paying attention or focusing on their studies, especially when the parent constantly interferes with the child's learning state.
- **Example**:  
  - Parent: "What are you daydreaming about? Focus and do your homework!"
  - Child: "I'm not daydreaming; I'm thinking about how to solve the problem!"

## Conflict Intensity Judgment Criteria

### Intensity Levels:
1. High 2. Medium 3. Low

### Judgment Criteria:

#### High
- **Tone**: The tone of the conversation is extremely intense, often involving shouting, criticism, or scolding.
- **Severity of Language**: The language includes insulting, belittling, or blaming remarks.
- **Length**: The conversation is lengthy, with escalating tension.
- **Body Language (if described)**: Involves intense gestures like slamming tables or throwing objects.

**Example**:
- Parent yells: "How can you be so useless? You can't even do this right!"
- Child shouts: "I've tried my best! Why do you always scold me?"

#### Medium
- **Tone**: The tone is relatively emotional but does not reach shouting, possibly involving strong dissatisfaction and dispute.
- **Severity of Language**: The language includes criticism or blame, but no serious insults.
- **Length**: The conversation is of medium length, with some dispute but not escalating further.

**Example**:
- Parent says: "Why did you do poorly again this time? It's so disappointing."
- Child replies: "I tried really hard; I'll do better next time."

#### Low
- **Tone**: The tone is calm or slightly dissatisfied, but generally controlled, possibly including mild criticism or suggestions.
- **Severity of Language**: The language does not contain severe negative words, focusing more on expressing disappointment or offering suggestions.
- **Length**: The conversation is short, with relatively mild conflict.

**Example**:
- Parent says: "This time's result isn't ideal; try harder next time."
- Child replies: "Okay, I'll work on it."

## Task Description
Using audio transcription text, identify **parent-child conflict** scenarios in conversations during homework assistance. Each scenario should be described with its trigger, process, specific behaviors of the parent and child, conflict type, and intensity level.

## Note: The number of conflicts in a single conversation is not fixed, and multiple conflicts may occur.

## Input Format
The input is a list of dictionaries, each representing a segment of transcribed text. Each dictionary contains the following keys:
- "id": the dialogue ID.
- 'speaker': the role of the speaker (parent, child, or others).
- "content": the string containing the transcribed text.

## Example Output
Below is an example of the output format, using JSON to describe **parent-child conflict** scenarios identified in homework assistance conversations. Each scenario includes the trigger, process, specific behaviors of the parent and child, conflict type, and intensity level. Additionally, the dialogue segment's start and end IDs corresponding to the conflict are marked.

```json
[
    {
        "scene_id": 1,
        "Start ID": The start ID of the dialogue segment corresponding to this conflict,
        "End ID": The end ID of the dialogue segment corresponding to this conflict,
        "trigger": "The child repeatedly made mistakes in math homework, and the parent became dissatisfied.",
        "process": "The parent noticed the child made a mistake in a simple addition problem and became increasingly impatient. The child showed dissatisfaction and began to resist.",
        "parent_behavior": "Parent criticizes: 'How could you get such an easy question wrong? I’ve already taught you!'",
        "child_behavior": "Child retorts: 'I’ve already done it several times! Why do you keep criticizing me?'",
        "conflict_type": "Expectation Conflict (EC)",
        "severity": "Medium"
    },
    {
        "scene_id": 2,
        "Start ID": The start ID of the dialogue segment corresponding to this conflict,
        "End ID": The end

 ID of the dialogue segment corresponding to this conflict,
        "trigger": "The child was distracted while writing pinyin, and after multiple reminders, the parent lost patience.",
        "process": "The parent noticed the child daydreaming during homework and, after several reminders, became strict. The child began to resist and refused to continue doing homework.",
        "parent_behavior": "Parent yells: 'What are you daydreaming about? Focus, or you won’t be allowed to watch TV after finishing your homework!'",
        "child_behavior": "Child mutters: 'I wasn’t daydreaming, I was thinking about how to write the problem!'",
        "conflict_type": "Focus Conflict (FC)",
        "severity": "High"
    },
    {
        "scene_id": 3,
        "Start ID": The start ID of the dialogue segment corresponding to this conflict,
        "End ID": The end ID of the dialogue segment corresponding to this conflict,
        "trigger": "The child insisted on using their own method during the learning process, and the parent tried to intervene.",
        "process": "The parent believed the child’s learning method was incorrect and tried to get the child to follow their instructions, but the child insisted on their own method, refusing to accept intervention.",
        "parent_behavior": "Parent says: 'You can’t do your homework like this. Look at the question carefully before you start!'",
        "child_behavior": "Child argues: 'I just want to do it this way. Why do I have to follow your instructions?'",
        "conflict_type": "Learning Method Conflict (LMC)",
        "severity": "Medium"
    },
    ...
]
```

## Output Format
Each scene should be described concisely with the trigger, process, specific behaviors of the parent and child, and the conflict type and intensity level. Additionally, mark the start and end IDs of the dialogue segment where the conflict occurred. Please strictly return the output in the following format:
```json
[
    {
        "scene_id": 1,
        "Start ID": The start ID of the dialogue segment corresponding to this conflict,
        "End ID": The end ID of the dialogue segment corresponding to this conflict,
        "trigger": "Trigger",
        "process": "Process",
        "parent_behavior": "Parent's Specific Behavior",
        "child_behavior": "Child's Specific Behavior",
        "conflict_type": "Conflict Type",
        "severity": "Intensity Level"
    }
    ...
]
```
\end{spverbatim}
\endgroup

\end{document}